\documentclass[12pt,a4paper]{article}

\usepackage{amsmath}          
\usepackage{amstext}          
\usepackage{amssymb}          
\usepackage{amsfonts}         
\usepackage{graphicx}         
\usepackage{pict2e}           
\usepackage{color}            
\usepackage{hyperref}

\renewcommand{\P}{\operatorname{\mathbb{P}}}
\renewcommand{\Re}{\operatorname{Re}}
\renewcommand{\Im}{\operatorname{Im}}
\newcommand{\sgn}{\operatorname{sgn}}
\newcommand{\tr}{\operatorname{tr}}
\newcommand{\Sp}{\operatorname{Sp}}
\newcommand{\Id}{\mathrm{Id}}
\newcommand{\C}{\mathbb{C}}
\newcommand{\Ch}{\widehat{\C}}
\newcommand{\R}{\mathcal{R}}

\newcommand{\dd}{\mathrm{d}}
\newcommand{\ee}{\mathrm{e}}
\newcommand{\ii}{\mathrm{i}}

\newcommand{\gen}{\mathrm{g}}
\newcommand{\Sin}{S_{\mathrm{in}}}
\newcommand{\Sout}{S_{\mathrm{out}}}
\newcommand{\wout}{w_{\mathrm{out}}}
\newcommand{\wRin}{w_{\mathrm{in}}^{\rm R}}

\definecolor{darkred}{rgb}{0.75,0,0}
\definecolor{darkgreen}{rgb}{0,0.75,0}
\definecolor{darkblue}{rgb}{0,0,0.75}

\newtheorem{lemma}{Lemma}
\newtheorem{prop}{Proposition}

\title{Probability of a single current}
\author{Sylvain Prolhac}
\date{%
	Laboratoire de Physique Th\'eorique, Universit\'e Toulouse III, France\\\quad	\\%
}

\begin{document}

\maketitle

\begin{abstract}
The Riemann surface associated with counting the current between two states of an underlying Markov process is hyperelliptic. We explore the consequences of this property for the time-dependent probability of that current for Markov processes with generic transition rates. When the system is prepared in its stationary state, the relevant meromorphic differential is in particular fully characterized by the precise identification of all its poles and zeroes.
\end{abstract}

\section{Introduction}
Given a Markov process modelling some physical system, one of the most basic questions of interest beyond the stationary measure is the statistics of probability currents between states. Stationary large deviations of those currents, which can be computed from the largest eigenvalue of an appropriate deformed generator $M(g)$ with the variable $g$ conjugate to the current, have received much attention. The full time-dependent transient statistics of the currents, which requires not only all the eigenvalues of the deformed generator but also the corresponding eigenvectors, is in general a more complicated problem.

Transient statistics and stationary large deviations are however not completely disconnected from each other: in many situations, analytic continuation from the groundstate of a parameter-dependent operator allows to recover the rest of the spectrum \cite{BW1969.1,DT1996.1}, in part or totality. This observation can be formalized in terms of the spectral curve $\det(\lambda\,\Id-M(g))=0$ with complex values of the parameter $g$ and the eigenvalue $\lambda$, and the corresponding compact Riemann surface $\R$. The branching structure of $\R$ in the variable $g$ leads in particular to exceptional points \cite{K1995.2}, where several eigenstates coincide, and which plays a key role for the physics of non-Hermitian systems \cite{H2012.1,AGU2020.1,BBK2021.1}.

For the statistics of currents, and more generally of integer counting processes $Q_{t}$ incremented at the transitions of an underlying Markov process, the probability can generally be expressed as the contour integral (\ref{P[int R]}) of a meromorphic differential on the Riemann surface $\R$ associated with the deformed generator $M(g)$ \cite{P2022.1}. The approach was in particularly used in the totally asymmetric simple exclusion process with periodic boundaries, where the Riemann surface is generated quite explicitly by two permutations specifying analytic continuation around $3$ branch points, and particularly simple expressions were found for the probability of the particle current with specific initial conditions \cite{P2020.2}.

The example where the process $Q_{t}$ counts the total number of transitions from a given state (labelled as $1$) to another state (labelled as $2$) for a general Markov process with a finite number of states was also considered in \cite{P2022.1}, section~3. The corresponding Riemann surface is simply the Riemann sphere $\Ch$ (i.e. the complex plane plus the point at infinity), of genus $\gen=0$. A rather explicit expression was in particular given for the probability of $Q_{t}$ when the Markov process is prepared with stationary initial condition. The main purpose of the present paper is to extend this simple example to a case with a non-trivial Riemann surface $\R\neq\Ch$, in order to understand the implications of a non-zero genus $\gen$ for the probability of $Q_{t}$ (the answer being essentially that contour integrals around poles are replaced here by contour integrals around branch cuts, and that $2\gen$ ``missing zeroes'' must be found).

The simple extension considered in this paper is the current $Q_{t}$ between states $1$ and $2$ (i.e. counting positively transitions from $1$ to $2$ and negatively transitions from $2$ to $1$) of a general Markov process with $\Omega\geq3$ states prepared initially in its stationary state. The associated Riemann surface then belongs to the simple family of hyperelliptic Riemann surfaces, generated by the square root of a polynomial $\Delta$ of degree $2\Omega$. This leads to a description of $\R$ in terms of two sheets that are exchanged when crossing either of the $\Omega$ branch cuts of $\sqrt{\Delta(\lambda)}$, leading to a genus $\gen=\Omega-1\geq2$, see figure~\ref{fig R5}. We emphasize that generic assumptions on the transition rates, ensuring in particular that the spectral curve is non-degenerate, i.e. the zeroes of $\Delta$ are distinct, are needed throughout the paper.

A key ingredient for the Riemann surface approach pursued here is the non-trivial factor, called $\mathcal{N}$, appearing in the integrand of the contour integral formula (\ref{P[int R]}) for the probability of the current $Q_{t}$. The quantity $\mathcal{N}$, defined in (\ref{N}), depends on the eigenvectors of $M(g)$ and on the initial condition of the Markov process. As a meromorphic function on a compact Riemann surface, $\mathcal{N}$ is in principle fully constrained by the knowledge of its poles and zeroes. As noted already in \cite{P2022.1} for a general integer counting process $Q_{t}$, the simplest case is when the Markov process is prepared in its stationary state, since this fixes some of the zeroes of $\mathcal{N}$, leaving $2\gen$ non-trivial zeroes, which are a priori unknown. In the simple example of the current between two states considered in this paper, these non-trivial zeroes are located precisely on $\R$ in lemma~\ref{lemma zeroes N 1} and \ref{lemma zeroes N 2}, which allows to reconstruct the function $\mathcal{N}$ and give an expression for the probability of $Q_{t}$.

It turns out that in the special case where the Markov process is reversible, the function $\mathcal{N}$ is symmetric under the exchange of the two sheets of the hyperelliptic Riemann surface $\R$, which leads to an especially simple expression for $\mathcal{N}$. The probability of $Q_{t}$ is then written as a real integral on the cuts of $\sqrt{\Delta(\lambda)}$, with rather explicit integrand (\ref{P[int cut] reversible}). This can be thought as the main result of the paper. It is illustrated in section~\ref{section 3 states model} on a simple $3$ states model.

When the Markov process is not reversible, the absence of symmetry between the two sheets of $\R$ leads for the function $\mathcal{N}$ to the less explicit expression (\ref{Mst[dlogMst]}), (\ref{dlogMst}), (\ref{M[N]}), written itself as the exponential of an integral. Additionally, $\mathcal{N}$ depends then on $\gen$ real constants $c_{\ell}$, that are to be fixed by a procedure outlined below (\ref{Mst[dlogMst]}). A non-reversible, almost symmetric and homogeneous one-dimensional random walk is worked out in section~\ref{section simple random walk} in order to illustrate in an explicit example that the constants $c_{\ell}$ are actually needed.

Here is the plan of the paper. In section~\ref{section Riemann surface}, we define the general process $Q_{t}$ considered in the paper (actually a slight generalization of the current between two states), discuss the various generic assumptions on the transition rates needed throughout the paper, and study the Riemann surface $\R$ and various special points on it needed later on. The statistics of $Q_{t}$ is then studied in section \ref{section probability}. After considering stationary large deviations, the general contour integral formula (\ref{P[int R]}) from \cite{P2022.1} is re-derived, and the poles and zeroes of the function $\mathcal{N}$ are located for the process $Q_{t}$ considered in this paper. The function $\mathcal{N}$ is finally reconstructed, first in the reversible case and then for a general non-reversible process.

\section{Hyperelliptic Riemann surface}
\label{section Riemann surface}
In this section, we define the process $Q_{t}$ counting the current between two states of a Markov process (and some extensions having the same analytic structure), and introduce the associated hyperelliptic Riemann surface.

\subsection{Markov counting processes considered}
We consider in this paper Markov processes on a finite number of states $j\in\{1,\ldots,\Omega\}$, $\Omega\geq3$. In terms of the $\Omega$-dimensional vector space spanned by the state vectors $|j\rangle$ and endowed with the scalar product $\langle k|j\rangle=\delta_{j,k}$, the probability vector $|P_{t}\rangle=\sum_{j}P_{t}(j)|j\rangle$, where $P_{t}(j)$ is the probability to observe the system at time $t$ in the state $j$, is solution of the master equation $\frac{\dd}{\dd t}|P_{t}\rangle=M|P_{t}\rangle$, with $M$ the $\Omega\times\Omega$ dimensional Markov operator. The transition rate from state $j$ to $k\neq j$ is denoted by $w_{k\leftarrow j}=\langle k|M|j\rangle$. At several points, generic assumptions will be needed for theses rates so as to prevent degeneracies for the Riemann surface and the meromorphic functions and differentials considered, see at the end of the section for more precise statements.

Conservation of probability at any time $t$, $\langle\Sigma|P_{t}\rangle=1$ with $\langle\Sigma|=\sum_{j=1}^{\Omega}\langle j|$, translates into $\langle\Sigma|M=0$ for the Markov operator, i.e. $\langle\Sigma|$ is a left eigenvector of $M$ with eigenvalue $0$. We assume ergodicity in the following, so that the eigenvalue $0$ is non-degenerate. The stationary state $|P_{\rm st}\rangle$ is then the unique right eigenvector of $M$ with eigenvalue $0$, and we write $P_{\rm st}(j)=\langle j|P_{\rm st}\rangle$ for the stationary probabilities.

We define also the reverse process, with transition rates
\begin{equation}
\label{wR}
w_{k\leftarrow j}^{\rm R}=w_{j\leftarrow k}\,\frac{P_{\rm st}(k)}{P_{\rm st}(j)}\;.
\end{equation}
The Markov matrix of the reverse process is $M^{\rm R}=D_{\rm st}M^{\top}D_{\rm st}^{-1}$ with $^{\top}$ indicating transposition and $D_{\rm st}$ the diagonal matrix $D_{\rm st}=\sum_{j=1}^{\Omega}P_{\rm st}(j)\,|j\rangle\langle j|$. The reverse process has the same stationary state as the original process, and $M^{\rm R}$ has the same eigenvalues as $M$.

If the Markov process is reversible, i.e. the product of transition rates for any cycle is equal to the product for the cycle with orientation reversed, the stationary state satisfies the detailed balance condition $w_{k\leftarrow j}P_{\rm st}(j)=w_{j\leftarrow k}P_{\rm st}(k)$ for any $j\neq k$, which allows to write the stationary probabilities as products of transition rates. Conjugation $D_{\rm st}^{-1/2}MD_{\rm st}^{1/2}$ then leads to a real symmetric matrix, and the spectrum of $M$ is real valued. Additionally, both original and reverse processes are identical in that case, i.e. $M^{\rm R}=M$. In the absence of reversibility, the stationary probabilities have more complicated expressions as sums over trees \cite{S1976.1}, the spectrum of $M$ may be complex valued, and $M^{\rm R}\neq M$.

\begin{table}
	\begin{center}
		\begin{tabular}{|c|c|c|}\hline
			&&\\[-4.5mm]
			Increments & Current $1\leftrightarrow2$ & General case\\\hline
			&&\\[-4.5mm]
			$Q_{t}\to Q_{t}+1$ & $1\to2$ & $1\to k$, $k\in\Sout$\\\hline
			&&\\[-4.5mm]
			$Q_{t}\to Q_{t}-1$ & $2\to1$ & $j\to1$, $j\in\Sin$\\\hline
		\end{tabular}\\[5mm]
		\begin{tabular}{|l|l|l|}\hline
			&&\\[-4mm]
			\begin{tabular}{c}
				Current $1\leftrightarrow2$\\[1mm]
				$\Sin=\Sout=\{2\}$\\[2mm]\\\quad
			\end{tabular}
			&
			\begin{tabular}{c}
				Reversible case\\[1mm]
				$\Sin=\Sout$\\[2mm]
				$w_{j\leftarrow k}^{\rm R}=w_{k\leftarrow j}$\\\quad
			\end{tabular}
			&
			\begin{tabular}{c}
				1d random walk\\[1mm]
				$\Sin=\Sout=\{2\}$\\[2mm]
				$w_{k\leftarrow j}=0$ if $|k-j|\neq1$\\
				\hspace*{26mm}$(\mathrm{mod}\;\Omega)$
			\end{tabular}\\\hline
		\end{tabular}
	\end{center}\vspace{-2mm}
	\caption{Counting processes $Q_{t}$ considered in this paper.}
	\label{table counting processes}
\end{table}

We are interested in the following in the probability current $Q_{t}$ between two states, say $1$ and $2$, defined as the number of transitions from $1$ to $2$ minus the the number of transitions from $2$ to $1$ between time $0$ and time $t$ (both $w_{2\leftarrow1}$ and $w_{1\leftarrow2}$ are assumed to be non-zero in the following). The random variable $Q_{t}$ is then an integer counting process \cite{HS2007.1}, and the statistics of $Q_{t}$ at a fixed time $t$ is characterized by the generating function
\begin{equation}
\label{GF[M(g)]}
\langle g^{Q_{t}}\rangle=\langle\Sigma|\ee^{tM(g)}|P_{0}\rangle\;,
\end{equation}
where the generator $M(g)$ is defined in terms of the Markov operator $M=M(1)$ as $\langle k|M(g)|j\rangle=g^{\delta_{j,1}\delta_{k,2}-\delta_{j,2}\delta_{k,1}}\,\langle k|M|j\rangle$.

Given two non-empty sets $\Sin,\Sout\subset\{2,\ldots,\Omega\}$ different from $\{2,\ldots,\Omega\}$, we consider more generally \footnote{In the following, $Q_{t}$ and $M(g)$ always refer to the counting process with arbitrary sets $\Sin$ and $\Sout$, unless explicitly stated otherwise.} the counting process $Q_{t}$ equal to the number of transitions from state $1$ to any state in $\Sout$ minus the number of transitions from any state in $\Sin$ to $1$, see table~\ref{table counting processes}. This process reduces to the current between the states $1$ and $2$ considered above when $\Sin=\Sout=\{2\}$. The identity (\ref{GF[M(g)]}) for the generating function of $Q_{t}$ is still satisfied, with a corresponding generator $M(g)$ defined as $\langle k|M(g)|j\rangle=g^{\sigma_{j,k}}\,\langle k|M|j\rangle$, where $\sigma_{j,k}=1$ if $j=1$ and $k\in\Sout$, $\sigma_{j,k}=-1$ if $j\in\Sin$ and $k=1$, and $\sigma_{j,k}=0$ otherwise.

Part of the analysis below (in particular the order of some poles) depends crucially on the form of the graph, i.e. on which rates are assumed to be equal to $0$. We always consider in the following $w_{k\leftarrow1}\neq0$ for $k\in\Sout$ and $w_{1\leftarrow j}\neq0$ for $j\in\Sin$. Other rates might be equal to zero, provided the assumptions stated at the end of the section are satisfied. A case with highly sparse Markov matrix where the main assumptions (A0), (A1), (A2) below are satisfied generically is the one-dimensional simple random walk with periodic boundary condition (simply referred to as the 1d random walk case in the following), where the only non-zero transition rates are $w_{k\leftarrow j}$ with $k=j\pm1$ (modulo $\Omega$), and the associated counting process $Q_{t}$ will always be chosen with $\Sin=\Sout=\{2\}$.

Associated with the counting process $Q_{t}$ with general sets $\Sin$, $\Sout$, we define the reverse counting process $Q_{t}^{\rm R}$ in terms of the generator
\begin{equation}
\label{MR(g)}
M^{\rm R}(g)=D_{\rm st}M(g)^{\top}D_{\rm st}^{-1}\;,
\end{equation}
such that $M^{\rm R}(1)=M^{\rm R}$ is the generator of the reverse Markov process. In the following, the counting process $Q_{t}$ is called \emph{reversible} if $M^{\rm R}(g)=M(g^{-1})$. The latter condition requires both that the underlying Markov process is reversible and that $\Sin=\Sout$, and implies in particular that the odd stationary cumulants of $Q_{t}$ vanish, see section~\ref{section stat LDF}, and additionally that $Q_{t}$ and $-Q_{t}$ have the same probability at any time $t$ if the system is prepared initially in its stationary state, see at the end of section~\ref{section exact formula Mst reversible}.

Quite a few assumptions are needed to carry out the analysis in the Riemann surface approach employed in this paper. These assumptions are however expected to hold for generic transition rates $w_{k\leftarrow j}$, and are relaxed somewhat by continuity for the probability of $Q_{t}$. Here is a list of the main assumptions used in the paper. Assumptions (A0), (A1), (A2) are always assumed from now on. Assumptions (A3), (A4), (A5), on the other hand, are stated where needed, and will be relaxed somewhat at the level of the final results.
\begin{itemize}
	\item (A0) $w_{k\leftarrow1}\neq0$ for $k\in\Sout$ and $w_{1\leftarrow j}\neq0$ for $j\in\Sin$, without which the process $Q_{t}$ does not make much sense.
	\item (A1) the eigenvalues of the Markov operator $M$ are distinct, which implies in particular the ergodicity of the Markov process and the uniqueness of the stationary state.
	\item (A2) the algebraic curve (\ref{spectral curve}) below is non-degenerate, i.e. the zeroes $\lambda_{i}$ of the polynomial $\Delta$ defined in (\ref{Delta}) below are all distinct. This implies in particular that both polynomials $P_{+}$ and $P_{-}$ defined in (\ref{det}) below do not vanish identically, otherwise $g$ would be a rational function of $\lambda$ and the corresponding Riemann surface would be the Riemann sphere and not the expected hyperelliptic Riemann surface. The assumption (A2) then implies that the spectrum of the deformed generator $M(g)$ depends on $g$, which is essential for the Riemann surface approach pursued in this paper, since otherwise the spectrum of $M(g)$ for all $g$ is just a fixed set of discrete points and does not generate a surface. In the reversible case, the assumption (A2) implies (A1), see section~\ref{section ramification lambda}.
	\item (A3) the polynomial $P_{+}$ (respectively $P_{-}$) has its maximal degree, specified in the first three rows of table~\ref{table deg P+-1}, and its zeroes are distinct. This assumption simplifies the counting of the points on the Riemann surface with $g=0$ and $g=\infty$.
	\item (A4)
	\begin{enumerate}
		\item the Markov operators $M_{\times}$ and $M_{\times}^{\rm R}$ of the modified Markov processes defined in section \ref{section non-trivial zeroes Nst} have non-degenerate eigenvalues (and additionally the non-zero eigenvalues of $M_{\times}$ and $M_{\times}^{\rm R}$ are distinct in the non-reversible case).
		\item the value $g_{*}$ of $g$ in lemma~\ref{lemma zeroes N 1} and \ref{lemma zeroes N 2} is finite, non-zero, and not a branch point for $g$.
		\item $g_{*}\neq1$ in lemma~\ref{lemma zeroes N 1} and \ref{lemma zeroes N 2}.
	\end{enumerate}
	The second part is needed in the proof of lemma~\ref{lemma zeroes N 1} and \ref{lemma zeroes N 2}, while the first and third part ensure that lemma~\ref{lemma zeroes N 1} and \ref{lemma zeroes N 2} give all the ``missing zeroes'' necessary to reconstruct the function $\mathcal{N}$ defined in (\ref{N}), with stationary initial condition.
	\item (A5) the various points on the Riemann surface appearing in table~\ref{table poles zeroes Nst} (or table~\ref{table poles zeroes Nst reversible} in the reversible case) are all distinct. Assumption (A5), which implies (A1), (A2), (A3) and (A4), is not strictly necessary but simplifies the reasoning in sections~\ref{section exact formula Mst reversible} and \ref{section exact formula Mst} for reconstructing the meromorphic function $\mathcal{N}$ with stationary initial condition from its poles and zeroes.
\end{itemize}

That the assumptions above are true for generic transition rates seems rather clear from the huge dependency on the transition rates of the polynomials involved in these assumptions, and is confirmed by numerics at least for smaller values of $\Omega$. Proper proofs might however not be straightforward for all the assumptions.

\begin{table}
	\begin{center}
		\begin{tabular}{|c|c|c|c|}\hline
			&&&\\[-4mm]
			Model & $\deg P_{0}$ & $\deg P_{+}$ & $\deg P_{-}$\\\hline
			&&\multicolumn{2}{c|}{}\\[-4mm]
			\begin{tabular}{c}
				Current $1\leftrightarrow2$\\
				generic rates\\
				$w_{k\leftarrow j}\neq0$
			\end{tabular}
			& $\Omega$&\multicolumn{2}{c|}{$\Omega-3$}\\[-4mm]
			&&\multicolumn{2}{c|}{}\\\hline
			&&&\\[-4mm]
			\begin{tabular}{c}
				General case\\
				generic rates\\
				$w_{k\leftarrow j}\neq0$
			\end{tabular}
			& $\Omega$
			& \begin{tabular}{c}$\Omega-3$ if $\Sout\subset\Sin$\\[2mm]$\Omega-2$ if $\Sout\not\subset\Sin$\end{tabular}
			& \begin{tabular}{c}$\Omega-3$ if $\Sin\subset\Sout$\\[2mm]$\Omega-2$ if $\Sin\not\subset\Sout$\end{tabular}\\[-4mm]
			&&&\\\hline
			&&\multicolumn{2}{c|}{}\\[-4mm]
			\begin{tabular}{c}
				Reversible case\\
				generic rates\\
				$w_{k\leftarrow j}=w_{j\leftarrow k}^{\rm R}\neq0$
			\end{tabular}
			& $\Omega$
			& \multicolumn{2}{c|}{\begin{tabular}{c}$\Omega-3$\\[2mm]($P_{+}=P_{-}$)\end{tabular}}\\[-4mm]
			&&\multicolumn{2}{c|}{}\\\hline
			&&\multicolumn{2}{c|}{}\\[-4mm]
			\begin{tabular}{c}
				$1$d random walk\\
				rates $w_{k\leftarrow j}\neq0$\\$|k-j|=1\mod\Omega$
			\end{tabular}
			& $\Omega$ & \multicolumn{2}{c|}{$0$}\\[-4mm]
			&&\multicolumn{2}{c|}{}\\\hline
		\end{tabular}
	\end{center}\vspace{-5mm}
	\caption{Degree of the polynomials $P_{0}$, $P_{+}$, $P_{-}$ in (\ref{det}) for the various counting processes considered, with generic transition rates $w_{k\leftarrow j}$.}
	\label{table deg P+-1}
\end{table}

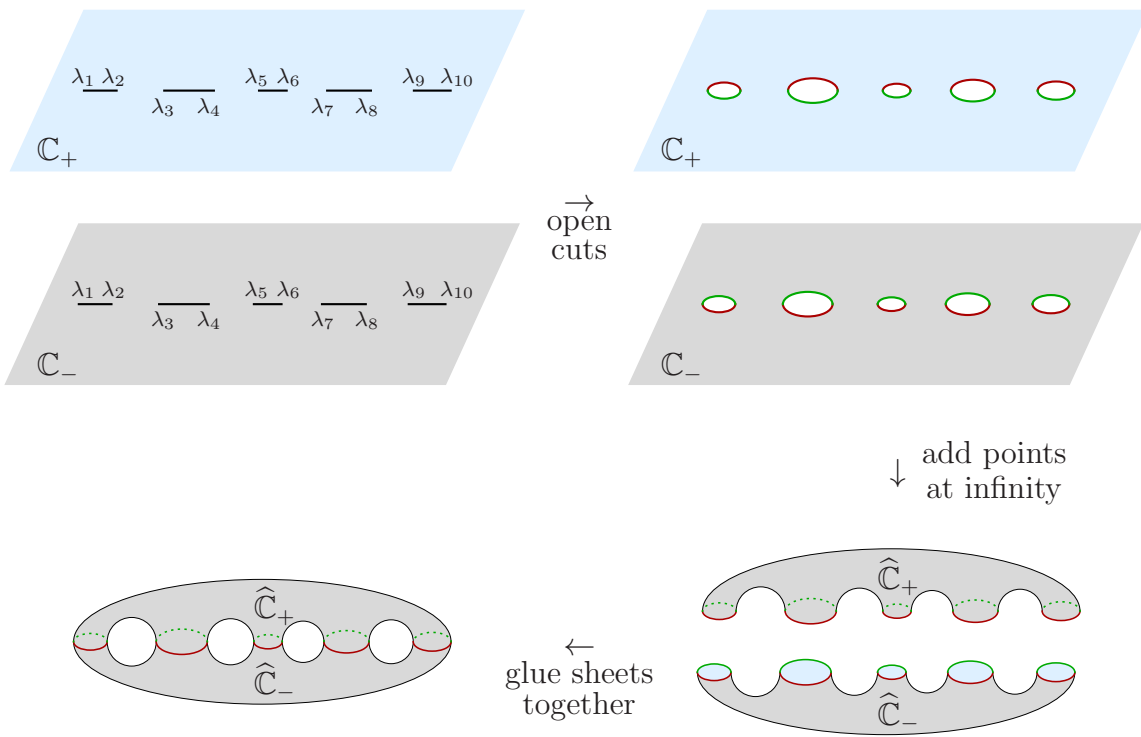
\begin{figure}
		\hspace*{-12mm}
		\begin{tabular}{ccc}
			\begin{tabular}{c}
				\includegraphics[width=68mm]{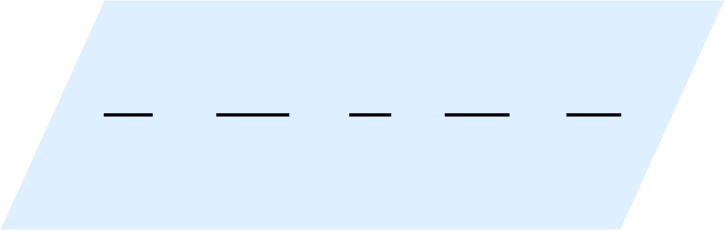}\\\\
				\includegraphics[width=68mm]{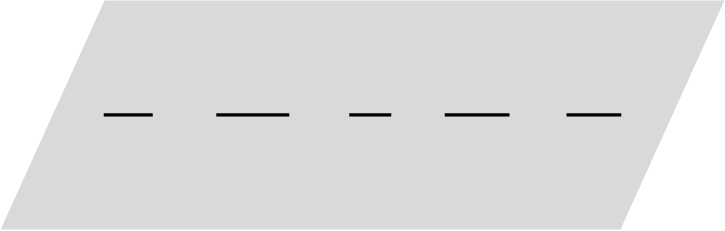}
				\begin{picture}(0,0)
					\put(-65,30){$\C_{+}$}\put(-65,1.5){$\C_{-}$}
					\put(-60.5,12){\scriptsize$\lambda_{1}$}\put(-56.5,12){\scriptsize$\lambda_{2}$}
					\put(-50,8){\scriptsize$\lambda_{3}$}\put(-44,8){\scriptsize$\lambda_{4}$}
					\put(-37.7,12){\scriptsize$\lambda_{5}$}\put(-33.5,12){\scriptsize$\lambda_{6}$}
					\put(-29,8){\scriptsize$\lambda_{7}$}\put(-23.2,8){\scriptsize$\lambda_{8}$}
					\put(-17,12){\scriptsize$\lambda_{9}$}\put(-12,12){\scriptsize$\lambda_{10}$}
					\put(-60.5,40.2){\scriptsize$\lambda_{1}$}\put(-56.5,40.2){\scriptsize$\lambda_{2}$}
					\put(-50,36.2){\scriptsize$\lambda_{3}$}\put(-44,36.2){\scriptsize$\lambda_{4}$}
					\put(-37.7,40.2){\scriptsize$\lambda_{5}$}\put(-33.5,40.2){\scriptsize$\lambda_{6}$}
					\put(-29,36.2){\scriptsize$\lambda_{7}$}\put(-23.2,36.2){\scriptsize$\lambda_{8}$}
					\put(-17,40.2){\scriptsize$\lambda_{9}$}\put(-12,40.2){\scriptsize$\lambda_{10}$}
				\end{picture}
			\end{tabular}
			&\hspace*{-30mm}$\underset{\displaystyle\text{open}\atop\displaystyle\text{cuts}}{\to}$\hspace*{-30mm}&
			\begin{tabular}{c}
				\includegraphics[width=68mm]{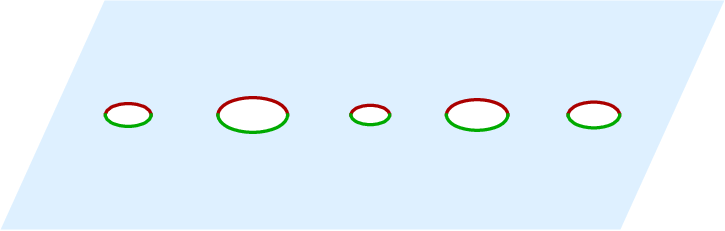}\\\\
				\includegraphics[width=68mm]{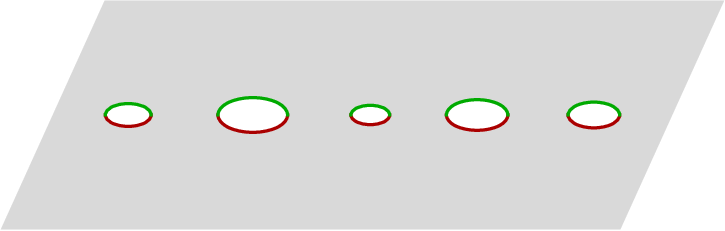}
				\begin{picture}(0,0)
					\put(-65,30){$\C_{+}$}\put(-65,1.5){$\C_{-}$}
				\end{picture}
			\end{tabular}\\\\
			&& \hspace*{25mm}$\downarrow$\begin{tabular}{c}add points\\[-0.5mm]at infinity\end{tabular}\\\\
			\begin{tabular}{c}
				\includegraphics[width=50mm]{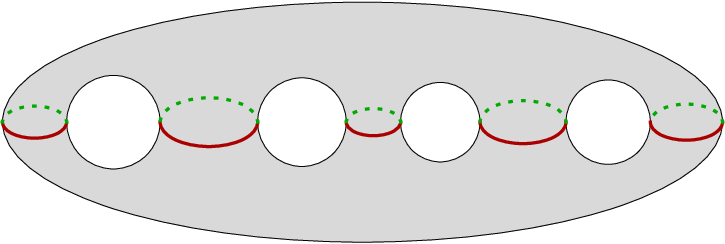}
				\begin{picture}(0,0)
					\put(-27.5,11.5){$\Ch_{+}$}\put(-27.5,1.5){$\Ch_{-}$}
				\end{picture}
			\end{tabular}
			&\hspace*{-30mm}$\underset{\displaystyle\text{glue sheets}\atop\displaystyle\text{together}}{\leftarrow}$\hspace*{-30mm}&
			\begin{tabular}{c}
				\includegraphics[width=50mm]{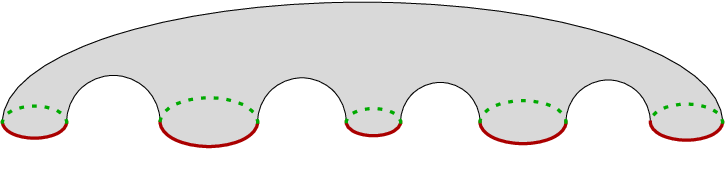}\\
				\includegraphics[width=50mm]{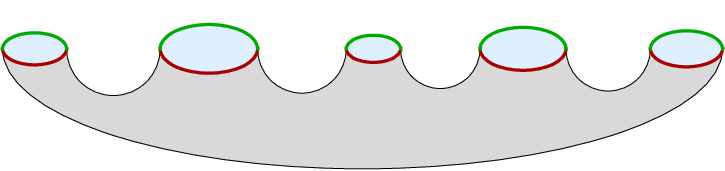}
				\begin{picture}(0,0)
					\put(-27.5,19.5){$\Ch_{+}$}\put(-27.5,1.5){$\Ch_{-}$}
				\end{picture}
			\end{tabular}
		\end{tabular}
	\caption{Riemann surface $\R$ of genus $\gen=4$ obtained by gluing together along the cuts (chosen as intervals of $\mathbb{R}$ for simplicity) the two sheets $\Ch_{+}$ and $\Ch_{-}$, for the counting process $Q_{t}$ considered in this paper, with $\Omega=5$ states.}
	\label{fig R5}
\end{figure}

\subsection{Riemann surface}
In this section, we consider the compact Riemann surface $\R$ associated with the counting process $Q_{t}$ defined in the previous section, and discuss some properties of meromorphic functions \footnote{A meromorphic function $f$ on the Riemann surface $\R$ is a function whose only singularities are poles. Extra branch points incompatible with $\R$ are not allowed for $f$, i.e. analytic continuation of $f$ along any closed path on $\R$ must not change the value of the function.	Branch points of $f$ in a given parametrization (e.g. $g$) of $\R$ are however allowed at the ramification points of $\R$ for that parametrization, as long as the change of variable to a proper local parameter of $\R$ around that point leads to an analytic behaviour for $f$.} on $\R$ corresponding to the parameter $g$ and the spectrum of $M(g)$. We refer to \cite{B2013.3,E2018.1,G2021.1} for detailed introductions about compact Riemann surfaces.

\subsubsection{Polynomials}
A crucial point for the following, which motivated our choice of counting process $Q_{t}$ in the previous section, is that the only entries of $M(g)$ actually depending on $g$ are in the first column and in the first row ($M(g)$ is actually a rank one perturbation of a constant Markov matrix $M_{\times}$, see section~\ref{section non-trivial zeroes Nst}). Since the entries of the first column (respectively first row) are either constant or proportional to $g$ (resp. $g^{-1}$), this implies
\begin{equation}
\label{det}
\det\big(\lambda\,\Id-M(g)\big)=P_{0}(\lambda)+gP_{+}(\lambda)+g^{-1}P_{-}(\lambda)\;,
\end{equation}
with $P_{0}$, $P_{+}$ and $P_{-}$ polynomials. The Riemann surface corresponding to $Q_{t}$ (which is the same as the one for the reverse process $Q_{t}^{\rm R}$) is then hyperelliptic, with a very simple branching structure in the variable $\lambda$, see the next section.

The polynomial $P_{0}$ in (\ref{det}) is always of degree $\Omega$, with leading coefficient equal to $1$. The degrees of the polynomials $P_{\pm}$ depends on the counting process considered and on the choice of transition rates, see table~\ref{table deg P+-1} for the examples mentioned above. One has always
\begin{equation}
\label{m+-}
m_{\pm}=\Omega-\deg P_{\pm}\geq2\;.
\end{equation}
In particular, $m_{+}=m_{-}$ generically when $\Sin=\Sout$, and furthermore $P_{+}=P_{-}$ if $Q_{t}$ is reversible. Additionally, in the special case of the one-dimensional random walk, $P_{+}=-\prod_{j=1}^{\Omega}w_{j+1\leftarrow j}$ and $P_{-}=-\prod_{j=1}^{\Omega}w_{j\leftarrow j+1}$ are constant, and $\det(\lambda\,\Id-M(g))$ is independent of the bond $k\to k+1$ across which the current $Q_{t}$ is counted.

The following lemma~\ref{lemma positivity 1}, which is a consequence of lemma~\ref{lemma positivity 2} below, fixes the signs of the coefficients of $P_{0}$ and $P_{\pm}$.

\begin{lemma}
\label{lemma positivity 1}
the coefficients of the polynomial $P_{0}$ are non-negative and the coefficients of the polynomials $P_{+}$ and $P_{-}$ are non-positive.
\end{lemma}

\begin{lemma}
\label{lemma positivity 2}
for any Markov matrix $M$ in continuous time (i.e. with non-negative non-diagonal elements, and each column summing to $0$), the determinant $\det(\lambda\,\Id-M)$ and any cofactor $c_{i,j}(\lambda\,\Id-M)$ (defined as $(-1)^{i+j}$ times the determinant of the matrix obtained by removing the $i$-th row and $j$-th column of $\lambda\,\Id-M$) are polynomials in $\lambda$ with non-negative coefficients.
\end{lemma}

\noindent
\textit{Proof of lemma \ref{lemma positivity 2}:} given a $\Omega\times\Omega$ matrix $A$, the main theorem in \cite{H1979.1} \footnote{With $A$ the transpose of the one in \cite{H1979.1}. Additionally, the positivity constraints on $A$ in the statement of the theorem in \cite{H1979.1} are not necessary (and not used in the proof there either). Indeed, the cofactors are obviously polynomials in the matrix elements of $A$.}, proved by induction on $\Omega$, states that any cofactor $c_{i,j}(\Id-A)$ can be written as a sum of terms, each term being a product of exactly $\Omega-1$ factors (without additional signs or numerical coefficient) taken among the matrix elements $a_{i,j}$ of $A$ and the coefficients $r_{j}=1-\sum_{i=1}^{\Omega}a_{i,j}$.\\
Considering then a general Markov matrix $M$ with size $\Omega\times\Omega$, and writing $M=N-D$ with $N$ and $D$ containing respectively the non-diagonal and diagonal elements of $M$, one has $\lambda\,\Id-M=(\lambda+\mu)\Id-N-E$ with $\mu\geq0$ the largest element of $D$, and $E=\mu\,\Id-D$. Then, one has $c_{i,j}(\lambda\,\Id-M)=(\lambda+\mu)^{\Omega-1}\,c_{i,j}(\Id-A)$ with $A=\frac{N+E}{\lambda+\mu}$. The theorem from \cite{H1979.1} stated above can then be applied to $c_{i,j}(\lambda\,\Id-M)/(\lambda+\mu)^{\Omega-1}$. With the notations above, $1-r_{j}=\frac{\lambda}{\lambda+\mu}$ since $M$ is a Markov matrix. Multiplying by $(\lambda+\mu)^{\Omega-1}$ then gives $c_{i,j}(\lambda\,\Id-M)$ as a sum of terms, where each term is the product of matrix elements of $N+E$ (which are non-negative since $M$ is a Markov matrix) times some non-negative powers of $\lambda$ and $\lambda+\mu$, and $c_{i,j}(\lambda\,\Id-M)$ is then a polynomial in $\lambda$ with non-negative coefficients. The case of the determinant $\det(\lambda\,\Id-M)$ can finally be understood e.g. as the cofactor $c_{1,1}$ for a matrix $M$ with an extra empty first row and first column.
\\\rightline{$\square$}\\

\noindent
\textit{Proof of lemma \ref{lemma positivity 1}:} from (\ref{det}) and the definition of the generator $M(g)$, the polynomials $P_{+}$ and $P_{-}$ can be written in terms of cofactors of $\lambda\,\Id-M$ with $M$ the (non-deformed) Markov operator as $P_{+}(\lambda)=-\sum_{k\in\Sout}w_{k\leftarrow1}\,c_{k,1}(\lambda\,\Id-M)$ and $P_{-}(\lambda)=-\sum_{j\in\Sin}w_{1\leftarrow j}\,c_{1,j}(\lambda\,\Id-M)$, and lemma \ref{lemma positivity 2} allows to conclude for $P_{\pm}$. Then, setting $g=1$ in (\ref{det}), one has $P_{0}(\lambda)=\det(\lambda\,\Id-M)-P_{+}(\lambda)-P_{-}(\lambda)$. Combining the result for $P_{\pm}$ and lemma \ref{lemma positivity 2} again, which ensures the positivity of the coefficients of $\det(\lambda\,\Id-M)$, finally allows to conclude for $P_{0}$.
\\\rightline{$\square$}\\

We finally introduce the polynomial
\begin{equation}
\label{Delta}
\Delta(\lambda)=P_{0}(\lambda)^{2}-4P_{+}(\lambda)P_{-}(\lambda)=\prod_{i=1}^{2\Omega}(\lambda-\lambda_{i})\;.
\end{equation}
The zeroes $\lambda_{i}$ of $\Delta$, which play an important role in the following as branch points in the variable $\lambda$, are generally complex numbers. The study of stationary large deviations in section~\ref{section stat LDF} shows that at least two of them must be real, and that all the real $\lambda_{i}$ are non-positive. Incidentally, numerics also suggest that $\Re\lambda_{i}\leq0$ for all $i$, although this property is not used in the following. Additionally, all the $\lambda_{i}$ are real numbers in the special case where $Q_{t}$ is reversible, and also for the 1d random walk, see section~\ref{section ramification lambda}.

\subsubsection{Algebraic curve}
The counting process $Q_{t}$ is associated with the complex algebraic curve
\begin{equation}
\label{spectral curve}
\det(\lambda\,\Id-M(g))=0\;,
\end{equation}
called the spectral curve of $Q_{t}$, whose solutions $(\lambda,g)\in\C^{2}$ are such that $\lambda$ is an eigenvalue of $M(g)$. Given a solution of (\ref{spectral curve}), (\ref{det}) leads to $g=\frac{-P_{0}(\lambda)\pm\sqrt{\Delta(\lambda)}}{2P_{+}(\lambda)}$. For each $\lambda$ not a zero of $\Delta$, there are exactly two distinct values of $g$ such that $\lambda$ is an eigenvalue of $M(g)$. Consider now the complex algebraic curve
\begin{equation}
\label{hyperelliptic curve}
y^{2}=\Delta(\lambda)\;.
\end{equation}
Under assumption (A2) that all the $2\Omega$ zeroes $\lambda_{i}$ of $\Delta$ are distinct, which is true if the transition rates $w_{k\leftarrow j}$ are generic, the algebraic curve (\ref{hyperelliptic curve}) is non-singular. We then call $\R$ the compact Riemann surface obtained after properly adding the points with $|\lambda|\to\infty$, see e.g. \cite{B2013.3}.

For $\Omega\geq3$ (the case $\Omega=2$ is not considered since it leads to $P_{+}=P_{-}=0$), the algebraic curve is called hyperelliptic, and has genus
\begin{equation}
\label{genus}
\gen=\Omega-1\;,
\end{equation}
see figure~\ref{fig R5}. The corresponding hyperelliptic Riemann surfaces constitute a special submanifold of complex dimension $2\Omega-3=2\gen-1$ ($\lambda_{i}\in\C$ modulo Möbius transformations) of the moduli space of compact Riemann surfaces of genus $\gen$, which has complex dimension $3\gen-3$. A nice feature of the hyperelliptic case is that meromorphic differentials with simple poles can be constructed quite explicitly.

\subsubsection{Ramification for \texorpdfstring{$\lambda$}{lambda}}
\label{section ramification lambda}
Taking the square root of (\ref{hyperelliptic curve}), any point of $\R$ can be labelled in a unique way as $[\lambda,\pm]$, $\lambda\in\Ch=\C\cup\{\infty\}$, except for the identification
\begin{equation}
p_{i}=[\lambda_{i},+]=[\lambda_{i},-]
\end{equation}
for the zeroes $\lambda_{i}$ of $\Delta$. This defines a splitting of $\R$ into two sheets $\Ch_{\pm}=\{[\lambda,\pm],\lambda\in\Ch\}$, whose intersection is made of the $2\Omega$ points $p_{i}$. We also write $\C_{\pm}$ for the corresponding copies of the complex plane, excluding the points at infinity $[\infty,\pm]$.

The precise definition of the sheets $\Ch_{\pm}$ depends crucially on the choice of square root for (\ref{hyperelliptic curve}). We group the zeroes $\lambda_{i}$ of $\Delta$ into pairs, say $(\lambda_{2i-1},\lambda_{2i})$, $i=1,\ldots,\Omega$ for some ordering of the $\lambda_{i}$, and choose a set of $\Omega$ non-intersecting paths $\gamma_{i}\subset\C$ between the $\lambda_{2i-1}$ and $\lambda_{2i}$. Then, we call $\sqrt{\Delta(\lambda)}$ the unique square root of $\Delta(\lambda)$ analytic outside the paths $\gamma_{i}$ such that $\sqrt{\Delta(\lambda)}\simeq+\lambda^{\Omega}$ when $\lambda\to\infty\in\Ch$. The definition may be extended to the cuts $\gamma_{i}$ by specifying for each one the side of the cut from which $\sqrt{\Delta(\lambda)}$ is continuous, which leads to a function $\sqrt{\Delta(\lambda)}$ defined for all $\lambda\in\Ch$, with branch points $\lambda_{i}$ and branch cuts $\gamma_{i}$ across which $\sqrt{\Delta(\lambda)}$ is discontinuous. The sheets $\Ch_{\pm}$ of $\R$, glued together along the cuts as in figure~\ref{fig R5}, are then made of the points of the algebraic curve (\ref{hyperelliptic curve}) with $y=\pm\sqrt{\Delta(\lambda)}$ for any $\lambda\in\Ch$.

Considering from now on $\lambda$ and $y$ as the meromorphic functions on $\R$ defined as
\begin{eqnarray}
\label{lambda(p)}
\lambda([\lambda_{*},\pm])&=&\lambda_{*}\\
\label{y(p)}
y([\lambda_{*},\pm])&=&\pm\sqrt{\Delta(\lambda_{*})}\;,
\end{eqnarray}
the corresponding meromorphic function $g$ is finally chosen as
\begin{equation}
\label{g(p)}
g(p)=\frac{y(p)-P_{0}(\lambda(p))}{2P_{+}(\lambda(p))}=-\frac{2P_{-}(\lambda(p))}{y(p)+P_{0}(\lambda(p))}\;,
\end{equation}
where the second equality follows from (\ref{Delta}). This leads for $y(p)$ to the useful symmetrized expression
\begin{equation}
\label{y[g,lambda] sym}
y(p)=g(p)P_{+}(\lambda(p))-g(p)^{-1}P_{-}(\lambda(p))\;,
\end{equation}
which is independent on the choice of cuts $\gamma_{i}$ unlike (\ref{y(p)}). The choice of sign in the definition (\ref{y(p)}) is arbitrary. Changing it simply amounts to a composition with the hyperelliptic involution $p\to\overline{p}$ on $\R$, defined as
\begin{equation}
\label{hyperelliptic involution}
\overline{[\lambda,\pm]}=[\lambda,\mp]\;,
\end{equation}
and such that $\lambda(\overline{p})=\lambda(p)$, $y(\overline{p})=-y(p)$ and $g(\overline{p})=\frac{P_{-}(\lambda(p))}{P_{+}(\lambda(p))}\,\frac{1}{g(p)}$. Additionally, since $\R$ is hyperelliptic, any meromorphic function on $\R$ can be written in a unique way in the canonical algebraic form
\begin{equation}
\label{canonical form mero function}
f(p)=R(\lambda(p))+\frac{S(\lambda(p))}{y(p)}\;,
\end{equation}
with $R$ and $S$ rational functions (i.e. ratios of polynomials).

In the special case of the 1d random walk, where $P_{\pm}$ are constant, $y(p_{i})=0$ combined with (\ref{y[g,lambda] sym}) implies that the ramification points $p_{i}$ have $g(p_{i})=\pm\sqrt{P_{-}/P_{+}}$, and the $\lambda_{i}$ are thus the eigenvalues of $M(\pm\sqrt{P_{-}/P_{+}})$. Conjugation $U^{-1}M(\pm\sqrt{P_{-}/P_{+}})U$ by the diagonal matrix $U$ with entries $\langle1|U|1\rangle=1$ and $\langle j|U|j\rangle=\sqrt{\prod_{i=j}^{\Omega}\frac{w_{i\leftarrow i+1}}{w_{i+1\leftarrow i}}}$ for $j\neq1$ leads to a real symmetric matrix, which implies that all $\lambda_{i}\in\mathbb{R}$ in that case.

If the counting process $Q_{t}$ is reversible, one has $P_{+}=P_{-}$, and $y(p_{i})=0$ combined with (\ref{y[g,lambda] sym}) implies $g(p_{i})=\pm1$. The $2\Omega$ branch points $\lambda_{i}$ are thus the eigenvalues of $M(1)$ and $M(-1)$, which are real valued (conjugation by $D_{\rm st}^{1/2}$ with $D_{\rm st}$ defined below (\ref{wR}) leads in both cases to a real symmetric matrix), and assumed to be distinct by assumption (A2). Since $M(1)$ is a Markov matrix, its largest eigenvalue is $0$, and $\det(\lambda\Id-M(1))=P_{0}(\lambda)+P_{+}(\lambda)+P_{-}(\lambda)>0$ for any $\lambda>0$. This implies $\det(\lambda\Id-M(-1))=P_{0}(\lambda)-P_{+}(\lambda)-P_{-}(\lambda)>-2P_{+}(\lambda)-2P_{-}(\lambda)>0$ for $\lambda>0$, where the last inequality comes from the fact that the coefficients of $P_{\pm}$ are non-positive. The eigenvalues of $M(-1)$ are then negative, and the $\lambda_{i}$ can be ordered as $\lambda_{1}<\ldots<\lambda_{2\Omega}=0$. This also follows from the study of stationary large deviations, see section~\ref{section stat LDF}. Additionally, choosing the intervals $\gamma_{i}=[\lambda_{2i-1},\lambda_{2i}]$ as cuts for $\sqrt{\Delta(\lambda)}$ (which corresponds to the choice $\sqrt{\Delta(\lambda)}=\prod_{i=1}^{\Omega}\big((\lambda-\lambda_{2i-1})\sqrt{\frac{\lambda-\lambda_{2i}}{\lambda-\lambda_{2i-1}}}\big)$, where the square root on the right side is the standard one, $\sqrt{r\ee^{\ii\theta}}=\sqrt{r}\,\ee^{\ii\theta/2}$ for $-\pi<\theta\leq\pi$), one has $\lambda\in\mathbb{R}$ and $y\in\ii\mathbb{R}$ on the cuts. Then (\ref{g(p)}), (\ref{hyperelliptic curve}), (\ref{Delta}) imply that in the variable $g$, the cuts simply correspond to $|g|=1$.

Degenerations of the hyperelliptic curve (\ref{hyperelliptic curve}) happen when the transition rates are chosen so that several $\lambda_{i}$ coincide. This happen in particular when all the transition rates $w_{k\leftarrow1}\to0$, $k\in\Sout$, which implies $P_{+}\to0$, or when $w_{1\leftarrow j}\to0$, $j\in\Sin$, which implies $P_{-}\to0$. In both cases, $\Delta(\lambda)=P_{0}(\lambda)^{2}$, the branch cuts become vanishingly small, and $\R$ splits into two disconnected Riemann spheres $\Ch_{\pm}$. Another example is when all the forward transition rates $w_{j+1\leftarrow j}\to w_{+}$ and backward transition rates $w_{j-1\leftarrow j}\to w_{-}$ for the 1d random walk, where (\ref{P0 random walk}) below implies that $\Delta(\lambda)/(\lambda^{2}+2(w_{+}+w_{-})\lambda+(w_{+}-w_{-})^{2})$ is again the square of a polynomial, and all the branch cuts except one disappear.

\begin{figure}
	\begin{center}
	\begin{picture}(80,40)
		{\color{black}
			\put(-12.5,0){\line(1,0){100}}\put(7.5,40){\line(1,0){100}}
			\put(-12.5,0){\line(1,2){20}}\put(87.5,0){\line(1,2){20}}
			\put(85,20){\circle*{1}}
			\put(84,21.5){$o$}
		}%
		{\color[rgb]{0.7,0.7,0.7}
			\multiput(80.5,20)(2,0){9}{\line(1,0){1}}
			\put(98,3){\small$\lambda_{\rm st}(\nu)$}
			\put(101,6){\vector(-2,3){8.5}}
			\put(101,1){\vector(-1,-2){11}}
		}%
		{\color{darkred}\thicklines
			\put(5,20){\line(1,0){10}}\put(8.5,17){$\gamma_{1}$}
			\qbezier(25,20)(45,20)(32.5,7.5)\put(32,15){$\gamma_{2}$}
			\qbezier(60,20)(55,32.5)(45,32.5)\put(51,26){$\gamma_{3}$}
			\put(65,20){\line(1,0){15}}\put(71,22){$\gamma_{4}$}
			\put(5,20){\circle*{1}}
			\put(15,20){\circle*{1}}
			\put(25,20){\circle*{1}}
			\put(32.5,7.5){\circle*{1}}
			\put(45,32.5){\circle*{1}}
			\put(60,20){\circle*{1}}
			\put(65,20){\circle*{1}}
			\put(80,20){\circle*{1}}
		}%
		{\color{darkblue}
			\put(20,30){$p_{\infty}^{+}$}\put(26,31.5){\vector(3,1){20}}\put(19,32.5){\vector(-1,1){6}}\put(18.5,29){\vector(-4,-1){16}}
			\put(63,33){\circle*{1}}\put(66,32){\small$[\lambda_{*},+]$, $P_{-}(\lambda_{*})=0$}
		}%
		\put(-8,2){$\C_{+}$}
	\end{picture}\\[2mm]
	\begin{picture}(80,40)
		{\color{black}
			\put(-12.5,0){\line(1,0){100}}\put(7.5,40){\line(1,0){100}}
			\put(-12.5,0){\line(1,2){20}}\put(87.5,0){\line(1,2){20}}
			\put(85,20){\circle*{1}}
			\put(84,21.5){$\overline{o}$}
		}%
		{\color[rgb]{0.7,0.7,0.7}
			\multiput(80.5,20)(2,0){9}{\line(1,0){1}}
		}%
		{\color{darkred}\thicklines
			\put(5,20){\line(1,0){10}}\put(8.5,17){$\gamma_{1}$}
			\qbezier(25,20)(45,20)(32.5,7.5)\put(32,15){$\gamma_{2}$}
			\qbezier(60,20)(55,32.5)(45,32.5)\put(51,26){$\gamma_{3}$}
			\put(65,20){\line(1,0){15}}\put(71,22){$\gamma_{4}$}
			\put(5,20){\circle*{1}}
			\put(15,20){\circle*{1}}
			\put(25,20){\circle*{1}}
			\put(32.5,7.5){\circle*{1}}
			\put(45,32.5){\circle*{1}}
			\put(60,20){\circle*{1}}
			\put(65,20){\circle*{1}}
			\put(80,20){\circle*{1}}
		}%
		{\color{darkgreen}
			\put(20,30){$p_{\infty}^{-}$}\put(26,31.5){\vector(3,1){20}}\put(19,32.5){\vector(-1,1){6}}\put(18.5,29){\vector(-4,-1){16}}
			\put(47,10){\circle*{1}}\put(50,9){\small$[\lambda_{*},-]$, $P_{+}(\lambda_{*})=0$}
		}%
		\put(-8,2){$\C_{-}$}
	\end{picture}
\end{center}\vspace{-5mm}
\caption{Graphical representation of the sheets $\C_{\pm}$ of $\R$ for $\Omega=4$, $m_{\pm}=1$. The thick, red curves are the branch cuts $\gamma_{i}$ along which the sheets are glued together. The points with $g=0$ and $g=\infty$ are represented respectively in blue and green.}
\label{fig POI}
\end{figure}
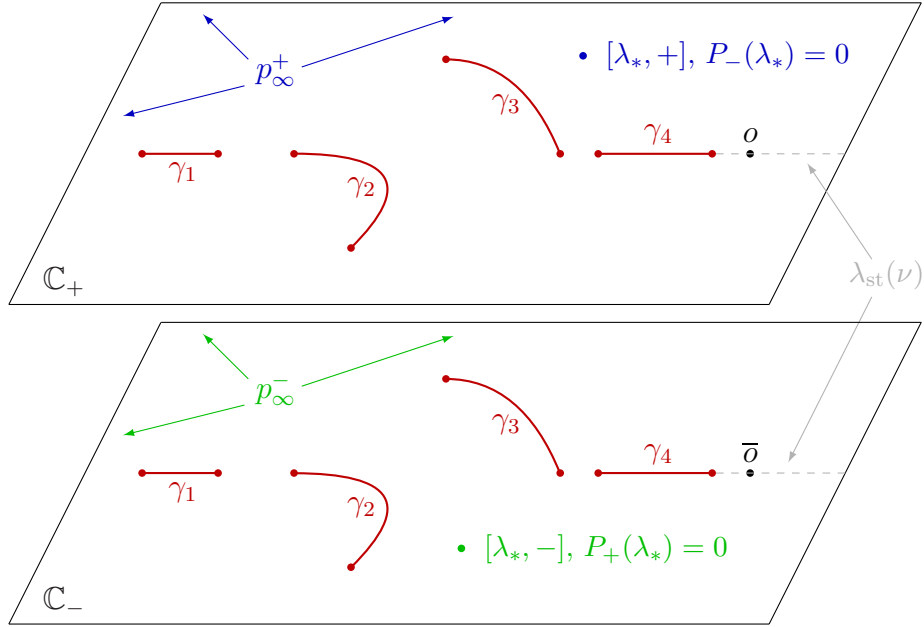

\begin{table}
	\begin{center}
		\begin{tabular}{|c|c|c|c|c|}\hline
			&&&&\\[-4.5mm]
			Point $p$ & Number & $y$ & $\lambda$ & $g$\\\hline
			&&&&\\[-4.5mm]
			\begin{tabular}{c}\color{darkred}{$p_{1}$, \ldots, $p_{2\Omega}$}\\(reversible: $g(p_{i})=\pm1$)\end{tabular} & $2\Omega$ & $1$ & $\cdot$ & $\cdot$\\\hline
			&&&&\\[-4.5mm]
			\begin{tabular}{c}$o$ and $\overline{o}$\\(reversible: {\color{darkred}$o=\overline{o}=p_{2\Omega}$})\end{tabular} & \begin{tabular}{c}$2$\\($1$)\end{tabular} & $\cdot$ & \begin{tabular}{c}$1$\\($2$)\end{tabular} & $\cdot$\\\hline
			&&&&\\[-4.5mm]
			\color{darkgreen}{$p_{\infty}^{-}=[\infty,-]$} & $1$ & $-\Omega$ & $-1$ & \begin{tabular}{r}$-m_{+}$\end{tabular}\\\hline
			&&&&\\[-4.5mm]
			\color{darkblue}{$p_{\infty}^{+}=[\infty,+]$} & $1$ & $-\Omega$ & $-1$ & \begin{tabular}{r}$m_{-}$\end{tabular}\\\hline
			&&&&\\[-4.5mm]
			\color{darkgreen}{\begin{tabular}{c}$[\lambda_{*},-]$,\\[1mm]$P_{+}(\lambda_{*})=0$\end{tabular}} & $\deg P_{+}$ & $\cdot$ & $\cdot$ & $-1$\\\hline
			&&&&\\[-4.5mm]
			\color{darkblue}{\begin{tabular}{c}{$[\lambda_{*},+]$,}\\[1mm]{$P_{-}(\lambda_{*})=0$}\end{tabular}} & $\deg P_{-}$ & $\cdot$ & $\cdot$ & $1$\\\hline
		\end{tabular}
	\end{center}\vspace{-5mm}
	\caption{Zeroes and poles of the functions $y$, $\lambda$ and $g$ on $\R$ for a counting process $Q_{t}$ satisfying assumptions (A1), (A2), (A3). Positive (respectively negative) numbers in the last three columns indicate the orders of the zeroes (resp. minus the orders of the poles). Dots corresponds to points that are neither poles nor zeroes. Comments and modifications in parentheses below refer to the reversible case. Colours for the points are the same as in figure~\ref{fig POI}.}
	\label{table POI}
\end{table}

\subsubsection{Points of interest on \texorpdfstring{$\R$}{R}}
The functions $\lambda:\R\to\Ch$, $g:\R\to\Ch$ and $y:\R\to\Ch$ are meromorphic. For generic $g_{*}\in\C$, $M(g_{*})$ has $\Omega$ distinct eigenstates corresponding to $\Omega$ points on $\R$, and $g$ has thus degree $\Omega$ (i.e. a generic value $g_{*}\in\C$ has $\Omega$ antecedents by $g$ on $\R$). From the discussion in the previous section, the function $\lambda$ on $\R$ has degree $2$. Finally, from (\ref{hyperelliptic curve}), to any value $y_{*}$ of $y$ correspond $\Omega$ values $\lambda_{*}$ of $\lambda$, and thus $2\Omega$ points on $\R$, and $y$ has degree $2\Omega$.

We consider in this section the poles and zeroes of $\lambda$, $g$ and $y$, see table~\ref{table POI} for a summary, which play a crucial role in sections~\ref{section exact formula Mst reversible} and \ref{section exact formula Mst} to identify the function $\mathcal{N}$ defined in (\ref{N}) below, with stationary initial condition. Around any point $p_{*}=[\lambda_{*},\pm]\in\R$, a proper local parameter $z$ (such that a small disk centred at $z=0$ corresponds to a small neighbourhood of $\R$ around $p_{*}$) must be chosen to identify the orders of the poles and zeroes. One can always choose $z=\lambda-\lambda_{*}$, except when $\Delta(\lambda_{*})=0$, where one can use instead $z=\sqrt{\lambda-\lambda_{*}}$ (or equivalently $z=y$), and when $\lambda_{*}=\infty$, where one can use $z=1/\lambda$.

From (\ref{y(p)}), (\ref{Delta}), $y$ has $2\Omega$ simple zeroes, the points $p_{i}=[\lambda_{i},+]=[\lambda_{i},-]$, which are ramified twice for $\lambda$. On the other hand, $y$ has only two poles,
\begin{equation}
\label{pinfty}
p_{\infty}^{\pm}=[\infty,\pm]\;,
\end{equation}
both of order $\Omega$.

When $0$ is not one of the $\lambda_{i}$, the function $\lambda$ has two simple zeroes. One of them, which we call the stationary point $o$, has $g(o)=1$ since $0$ is an eigenvalue of $M=M(1)$. From (\ref{g(p)}), (\ref{Delta}), the other zero $\overline{o}$ of $\lambda$ has $g(\overline{o})=P_{-}(0)/P_{+}(0)>0$. The respective sheets on which $o$ and $\overline{o}$ belong to depend on the sign of the stationary average $J$ of the counting process $Q_{t}$ as
\begin{equation}
o=[0,\sgn J]
\qquad\text{and}\qquad
\overline{o}=[0,-\sgn J]
\end{equation}
with $\sgn J=\sgn(P_{-}(0)-P_{+}(0))=\sgn(|P_{+}(0)|-|P_{-}(0)|)$, see section~\ref{section stat LDF} below. If $Q_{t}$ is reversible, $J=0$ and $o=\overline{o}=p_{2\Omega}$ is a pole of order $2$ of $\lambda$, ramified twice for $\lambda$. Furthermore, the function $\lambda$ has two simple poles, $p_{\infty}^{\pm}$ defined in (\ref{pinfty}).

Finally, we consider the poles and zeroes of $g$. Since by convention $y([\lambda,\pm])\simeq\pm\lambda^{\Omega}$ for large $\lambda$, (\ref{g(p)}) with $\deg P_{\pm}<\deg P_{0}$ leads to
\begin{equation}
g(p_{\infty}^{-})=\infty
\qquad\text{and}\qquad
g(p_{\infty}^{+})=0\;.
\end{equation}
Using (\ref{g(p)}), (\ref{m+-}) and calling $-c_{\pm}$, $c_{\pm}>0$ the leading coefficient of $P_{\pm}$, we observe more precisely that
\begin{equation}
\label{g large lambda}
g([\lambda,-])\simeq\frac{\lambda^{m_{+}}}{c_{+}}
\qquad\text{and}\qquad
g([\lambda,+])\simeq\frac{c_{-}}{\lambda^{m_{-}}}\;,
\end{equation}
which means that $g$ has a pole of order $m_{+}$ at $p_{\infty}^{-}$ and a zero of order $m_{-}$ at $p_{\infty}^{+}$. The missing poles (respectively zeroes) of $g$ are of the form $[\lambda_{*},\pm]$ with $\lambda_{*}$ a zero of $P_{+}$ (resp. $P_{-}$) and the sign such that $y([\lambda_{*},\pm])=-P_{0}(\lambda_{*})$ (resp. $y([\lambda_{*},\pm])=P_{0}(\lambda_{*})$). Incidentally, since if the counting process $Q_{t}$ is reversible, our conventions in section~\ref{section ramification lambda} impose that the branch cuts of $\sqrt{\Delta(\lambda)}$ are the curves $|g|=1$, one has $g^{-1}(\infty)\subset\Ch_{-}$ and $g^{-1}(0)\subset\Ch_{+}$ in that case. This is not true in general in the absence of reversibility, depending on the choice of cuts $\gamma_{i}$. However, deforming a cut across any $\lambda_{*}\in\C$ exchanges the points $[\lambda_{*},\pm]$, and it is thus always possible to choose the cuts $\gamma_{i}$ so that $g^{-1}(\infty)\subset\Ch_{-}$ and $g^{-1}(0)\subset\Ch_{+}$ as in the reversible case. We always adopt this convention in the following.

\subsubsection{Ramification for \texorpdfstring{$g$}{g}}
\label{section ramification g}
For the sake of completeness, we discuss in this section ramification of $\R$ in the variable $g$. The results are not directly used for the statistics of $Q_{t}$.

Ramification points $p_{*}\in\R$ for $g$, with $\lambda(p_{*})=\lambda_{*}$ and $g(p_{*})=g_{*}$ (also called exceptional points \cite{K1995.2}, and where several eigenstates of $M(g_{*})$ coincide), lie at the intersection of the complex algebraic curves $P(\lambda,g)=0$ and $\partial_{\lambda}P(\lambda,g)=0$, with $P(\lambda,g)=\det(\lambda\Id-M(g))$. Eliminating e.g. the term proportional to $g^{-1}$ between both equations leads to a rational expression for $g_{*}$ as a function of $\lambda_{*}$. Plugging this expression into $P(\lambda_{*},g_{*})=0$ gives an equation for $\lambda_{*}$ only,
\begin{equation}
\label{ramification points g}
\frac{P_{0}(\lambda_{*})P_{-}'(\lambda_{*})-P_{0}'(\lambda_{*})P_{-}(\lambda_{*})}{P_{+}(\lambda_{*})P_{-}'(\lambda_{*})-P_{+}'(\lambda_{*})P_{-}(\lambda_{*})}=\frac{P_{-}(\lambda_{*})P_{+}'(\lambda_{*})-P_{-}'(\lambda_{*})P_{+}(\lambda_{*})}{P_{0}(\lambda_{*})P_{+}'(\lambda_{*})-P_{0}'(\lambda_{*})P_{+}(\lambda_{*})}\;,
\end{equation}
which has $4\Omega-m_{+}-m_{-}-2$ solutions $\lambda_{*}\in\C$. For such $\lambda_{*}$, generically only one of the two points $[\lambda_{*},\pm]$ is ramified for $g$. On the other hand, when $Q_{t}$ is reversible, (\ref{ramification points g}) reduces to $(P_{0}(\lambda_{*})P_{+}'(\lambda_{*})-P_{0}'(\lambda_{*})P_{+}(\lambda_{*}))^{2}=0$ using $P_{+}=P_{-}$, and both points $[\lambda_{*},\pm]$ are ramified for $g$. This is also the case for the 1d random walk, where $P_{0}'(\lambda_{*})^{2}=0$ since $P_{+}$ and $P_{-}$ are constant.

From (\ref{g large lambda}), the points $p_{\infty}^{\pm}$ are also ramified $m_{\mp}$ times for $g$. The Riemann-Hurwitz formula for the genus $\gen$ of $\R$ (see e.g. \cite{B2013.3}),
\begin{eqnarray}
&\gen&=-d_{g}+1+\frac{1}{2}\sum_{p\in\R}\big(e_{g}(p)-1\big)\\
&&=-\Omega+1+\frac{(m_{+}-1)+(m_{-}-1)+(4\Omega-m_{+}-m_{-}-2)\times1}{2}\nonumber\\
&&=\Omega-1\nonumber
\end{eqnarray}
with $d_{g}=\Omega$ the degree of $g$ and $e_{g}(p)$ ramification indices for $g$, is indeed satisfied.

Ramification points for $\lambda$ and $g$ with a finite non-zero value for $g$ respectively appear as $2\Omega$ simple poles and $4\Omega-m_{+}-m_{-}-2$ simple zeroes of the meromorphic function $\frac{\dd\log g}{\dd\lambda}$, since corresponding ramification indices are equal to $2$ for $\lambda$, and also for $g$ generically. From (\ref{g large lambda}), the points $p_{\infty}^{\pm}$ are simple zeroes of $\frac{\dd\log g}{\dd\lambda}$. Additionally, the $\Omega-m_{+}$ (respectively $\Omega-m_{-}$) points $p\in g^{-1}(0)\setminus\{p_{\infty}^{+}\}$ (resp. $p\in g^{-1}(\infty)\setminus\{p_{\infty}^{-}\}$), corresponding to $\lambda(p)$ zero of $P_{+}$ (resp. $P_{-}$), are simple poles of $\frac{\dd\log g}{\dd\lambda}$ generically. The degree of $\frac{\dd\log g}{\dd\lambda}$ is then equal to $(2\Omega)+(\Omega-m_{+})+(\Omega-m_{-})=(4\Omega-m_{+}-m_{-}-2)+1+1$, and all the poles and zeroes of $\frac{\dd\log g}{\dd\lambda}$ are accounted for.

\section{Statistics of the current}
\label{section probability}
In this section, we use the Riemann surface defined in the previous section to express the probability of $Q_{t}$ at arbitrary time $t$, for a system prepared initially in its stationary state.

\subsection{Stationary large deviations}
\label{section stat LDF}
In the long time limit, the cumulant generating function $\log\langle\ee^{\nu Q_{t}}\rangle$ of $Q_{t}$ has for $\nu\in\mathbb{R}$ (i.e. $g=\ee^{\nu}>0$) the asymptotics
\begin{equation}
\log\langle\ee^{\nu Q_{t}}\rangle\simeq t\lambda_{\rm st}(\nu)\;,
\end{equation}
where $\lambda_{\rm st}(\nu)\in\mathbb{R}$ is the eigenvalue of $M(\ee^{\nu})$ with largest real part. The stationary cumulants $\langle Q_{t}^{n}\rangle_{\rm c}=\partial_{\nu}^{n}\log\langle\ee^{\nu Q_{t}}\rangle_{|\nu\to0}$ of $Q_{t}$ can be extracted by an expansion of the eigenvalue $\lambda$ around the stationary point $o\in\R$, as \footnote{Note that derivatives in (\ref{cumulants}) are with respect to the variable $\nu=\log g$ conjugate to $Q_{t}$, while derivatives in (\ref{J[P]}), (\ref{D[P]}) are with respect to the eigenvalue $\lambda$.}
\begin{equation}
\label{cumulants}
\lim_{t\to\infty}\frac{\langle Q_{t}^{n}\rangle_{\rm c}}{t}=\lambda_{\rm st}^{(n)}(0)
\end{equation}
From (\ref{hyperelliptic curve}), (\ref{det}), the first two stationary cumulants $J=\lambda_{\rm st}'(0)=\lim_{t\to\infty}\langle Q_{t}\rangle/t$ and $D=\lambda_{\rm st}''(0)=\lim_{t\to\infty}(\langle Q_{t}^{2}\rangle-\langle Q_{t}\rangle^{2})/t$ are in particular given by
\begin{equation}
\label{J[P]}
J=\frac{P_{-}(0)-P_{+}(0)}{P_{0}'(0)+P_{+}'(0)+P_{-}'(0)}
\end{equation}
and
\begin{equation}
\label{D[P]}
D=\frac{P_{0}(0)+2J(P_{-}'(0)-P_{+}'(0))-J^{2}(P_{0}''(0)+P_{+}''(0)+P_{-}''(0))}{P_{0}'(0)+P_{+}'(0)+P_{-}'(0)}\;.
\end{equation}

From (\ref{g large lambda}), one has
\begin{equation}
\lambda_{\rm st}(\nu)\simeq c_{\pm}^{1/m_{\pm}}\ee^{\pm\nu/m_{\pm}}\;\;\text{when $\nu\to\pm\infty$}\;,
\end{equation}
so that $\lambda_{\rm st}(\nu)\to+\infty$ when $\nu\to\pm\infty$. Additionally, since $\lambda:\R\to\Ch$ is of degree $2$, the function $\lambda_{\rm st}$ is strictly convex, as expected on general grounds for such large deviations \cite{T2009.1}.

Since $\lambda_{\rm st}(0)=0$, the function $\lambda_{\rm st}$ has then two distinct zeroes if $J\neq0$, corresponding to the points $o=[0,\sgn J]$ and $\overline{o}=[0,-\sgn J]$. The minimum $\lambda_{\rm st}^{\rm min}$ of $\lambda_{\rm st}$ corresponds to a branch point for $\lambda$, and is thus equal to the largest real zero of $\Delta$. Defining
\begin{equation}
\nu_{*}=\frac{1}{2}\log\Big(\frac{P_{-}(\lambda_{\rm st}^{\rm min})}{P_{+}(\lambda_{\rm st}^{\rm min})}\Big)\in\mathbb{R}
\end{equation}
and assuming that the cuts $\gamma_{i}$ are chosen so that they do not intersect the half-line $[\lambda_{\rm st}^{\rm min},\infty)$, the values of $\lambda_{\rm st}(\nu)$ for $\nu\leq\nu_{*}$ and $\nu\geq\nu_{*}$ correspond to points on distinct sheets $\C_{\pm}$ related by the hyperelliptic involution, and analytic continuation is needed in order to define $\lambda_{\rm st}(\nu)$ for all $\nu\in\mathbb{R}$, see figure~\ref{fig POI}. In the special case of the 1d random walk, where $P_{+}$ and $P_{-}$ are constant and $\nu_{*}=\frac{1}{2}\log(\prod_{j=1}^{\Omega}\frac{w_{j\leftarrow j+1}}{w_{j+1\leftarrow j}})$, this is simply the Gallavotti-Cohen symmetry $\lambda_{\rm st}(\nu)=\lambda_{\rm st}(2\nu_{*}-\nu)$ \cite{LS1999.1}.

When $J=0$, on the other hand, the point $o=\overline{o}$ is ramified twice for $\lambda$, which implies $\lambda_{\rm st}^{\rm min}=0$ and $\nu_{*}=0$. This happens in particular when $Q_{t}$ is reversible. In that case, the hyperelliptic involution (\ref{hyperelliptic involution}) simply exchanges $\nu$ with $-\nu$, so that $\lambda_{\rm st}$ is an even function and all the cumulants of $Q_{t}$ with odd order vanish.

Beyond stationary large deviations, expanding (\ref{GF[M(g)]}) over the eigenstates of $M(g)$, we observe that the cumulants of $Q_{t}$ behave as $\langle Q_{t}^{n}\rangle_{\rm c}\simeq t\lambda_{\rm st}^{(n)}(0)+\mu_{n}$, up to exponentially small corrections at late times, where the $\mu_{n}$ depend on the initial condition, unlike the leading coefficients $\lambda_{\rm st}^{(n)}(0)$. For stationary initial condition, expanding the left and right eigenvectors corresponding to $\lambda_{\rm st}(\nu)$ at first order near $\nu=0$, one finds in particular
\begin{equation}
\label{d1 st}
\mu_{1}^{\rm st}=0\;,
\end{equation}
i.e. $\langle Q_{t}\rangle-Jt$ vanishes exponentially fast at late times. In the reversible case, the coefficient $\mu_{2}$ is computed in (\ref{Var(Qt) late times}) below.

\subsection{Probability as a contour integral on $\R$}
The probability of $Q_{t}$ with general initial condition $|P_{0}\rangle$ can be extracted from (\ref{GF[M(g)]}) by residues as \footnote{We use the notation $h$ instead of $g$ here in order to avoid the confusion with the function $g$ on $\R$.}
\begin{equation}
\P(Q_{t}=Q)=\frac{1}{2\ii\pi}\oint_{\gamma}\frac{\dd h}{h^{Q+1}}\,\langle\Sigma|\ee^{tM(h)}|P_{0}\rangle\;,
\end{equation}
with $\gamma$ a simple closed path with positive orientation encircling $0$. For any $h\in\Ch$ not a branch point for $g$, expanding over the eigenstates of $M(h)$ amounts to summing over the antecedents of $h$ by the function $g$ on $\R$. One has $\langle\Sigma|\ee^{tM(h)}|P_{0}\rangle=\sum_{p\in g^{-1}(h)}\mathcal{N}(p)\,\ee^{t\lambda(p)}$ with $\mathcal{N}:\R\to\Ch$ meromorphic defined by
\begin{equation}
\label{N}
\mathcal{N}(p)=\frac{\langle\Sigma|\psi(p)\rangle\,\langle\psi(p)|P_{0}\rangle}{\langle\psi(p)|\psi(p)\rangle}\;.
\end{equation}
The left and right eigenvectors $\langle\psi(p)|$ and $|\psi(p)\rangle$ of $M(g(p))$ are chosen so that all their entries are meromorphic on $\R$. This can always be done since from $M(g)|\psi\rangle=\lambda|\psi\rangle$, the entries of the eigenvectors can be taken as rational functions of $\lambda$ and $g$. At the stationary point $o$, one has in particular
\begin{equation}
\label{N(o)}
\mathcal{N}(o)=1
\end{equation}
since $\langle\psi(o)|\propto\langle\Sigma|$ and $\langle\Sigma|P_{0}\rangle=1$.

Lifting $\gamma$ to each of the $\Omega$ sheets for $g$, the contour integral over $h\in\gamma$ of the sum over $p\in g^{-1}(h)$ above can be replaced by a contour integral on the union of closed curves $g^{-1}(\gamma)\subset\R$, which can be deformed into a simple closed curve $\Gamma\subset\R$ enclosing (respectively excluding) with positive orientation the points in $g^{-1}(0)$ (resp. $g^{-1}(\infty)$). Introducing
\begin{equation}
\label{M[N]}
\mathcal{M}=\frac{\dd\log g}{\dd\lambda}\,\mathcal{N}\;,
\end{equation}
and making the change of parametrization $\log g\to\lambda$ for $\Gamma$, we finally obtain
\begin{equation}
\label{P[int R]}
\P(Q_{t}=Q)=\frac{1}{2\ii\pi}\oint_{p\in\Gamma}\!\!\dd\lambda(p)\,\frac{\ee^{t\lambda(p)}\mathcal{M}(p)}{g(p)^{Q}}\;.
\end{equation}
With the convention that the cuts $\gamma_{i}$ are chosen so that $g^{-1}(0)\subset\C_{+}$ and $g^{-1}(\infty)\subset\C_{-}$, one can take for $\Gamma$ a closed curve with \emph{negative orientation} in the complex plane $\C_{+}$, enclosing all the cuts $\gamma_{i}$ and excluding the points in $g^{-1}(0)$, or a closed curve with \emph{positive orientation} in the complex plane $\C_{-}$, enclosing all the cuts $\gamma_{i}$ and excluding the points in $g^{-1}(\infty)$.

\subsection{Poles and zeroes of \texorpdfstring{$\mathcal{N}$}{N}, stationary initial condition}
Orthogonality of the eigenstates of $M(g)$ implies that the poles of $\mathcal{N}$ are located at ramification points $p$ for $g$, where a number of eigenstates equal to the ramification index $e_{g}(p)$ coincide. The corresponding order for the pole of $\mathcal{N}$ is equal to $e_{g}(p)-1$. Applying the Riemann-Hurwitz formula for $g$, the total number of poles of $\mathcal{N}$, counted with multiplicity, is then equal to $2\gen+2\Omega-2$ (i.e. $\mathcal{N}$ has degree $4\gen$ for the counting process considered in this paper, with arbitrary initial condition).

Since $\langle\Sigma|$ is an eigenvector of $M(1)$ with eigenvalue $0$, orthogonality of the eigenstates of $M(1)$ implies also that the $\Omega-1$ points in $g^{-1}(1)\setminus\{o\}$ are simple zeroes of $\mathcal{N}$ for generic initial condition, and even double zeroes of $\mathcal{N}$ for stationary initial condition since $|P_{\rm st}\rangle$ is also an eigenvector of $M(1)$ with eigenvalue $0$.

Since the meromorphic function $\mathcal{N}$ has the same number of poles and zeroes, equal to the degree of $\mathcal{N}$, some zeroes of $\mathcal{N}$ located outside $g^{-1}(1)$ are missing from the account above. We call them the non-trivial zeroes of $\mathcal{N}$. We specialize in the following to stationary initial condition and write $\mathcal{N}_{\rm st}$ for $\mathcal{N}$, which has then only $2\gen$ non-trivial zeroes. Lemmas~\ref{lemma zeroes N 1} and \ref{lemma zeroes N 2} of the next section give their locations on $\R$.

We emphasize that the counting above for the non-trivial zeroes of $\mathcal{N}_{\rm st}$ does not use the special form of $M(g)$ considered in this paper, and the number of non-trivial zeroes of $\mathcal{N}_{\rm st}$ is always equal to $2\gen$ for an integer counting process with generic transition rates \cite{P2022.1}.

In the hyperelliptic case considered in this paper, ramification points for $\lambda$ are simply the zeroes of $y$, while ramification points for $g$, which appear as poles of $\mathcal{N}_{\rm st}$, have a more complicated characterization through (\ref{ramification points g}). From the discussion at the end of section~\ref{section ramification g}, however, ramification points for $\lambda$ and $g$ appear as poles and zeroes of $\frac{\dd\log g}{\dd\lambda}$, and the ramification points for $g$ are in particular regular points for the product $\mathcal{M}_{\rm st}$ defined in (\ref{M[N]}), see table~\ref{table poles zeroes Nst}. Thus, since $\mathcal{M}_{\rm st}$ and $\mathcal{N}_{\rm st}$ have the same non-trivial zeroes, we work rather with $\mathcal{M}_{\rm st}$ in sections~\ref{section exact formula Mst reversible} and \ref{section exact formula Mst}.

\begin{table}
	\begin{center}
		\begin{tabular}{|c|c||c|c||c|c|c|}\hline
			&&&&&&\\[-4.5mm]
			Point $p$ & Number & $e_{g}$ & $e_{\lambda}$ & $\frac{\dd\log g}{\dd\lambda}$ & $\mathcal{M}_{\rm st}$ & $\mathcal{N}_{\rm st}$\\\hline
			&&&&&&\\[-4.5mm]
			\begin{tabular}{c}Ramification\\points for $\lambda$\\\color{darkred}$p_{1}$, \ldots, $p_{2\Omega}$\end{tabular} & $2\Omega$ & $1$ & $2$ & $-1$ & $-1$ & $0$\\\hline
			&&&&&&\\[-4.5mm]
			\begin{tabular}{c}Ramification\\points for $g$\\$[\lambda_{*},+]$ or $[\lambda_{*},-]$\\$\lambda_{*}\neq\infty$ as in (\ref{ramification points g})\end{tabular} & \begin{tabular}{l}$4\Omega-m_{+}$\\\;$-m_{-}-2$\end{tabular} & $2$ & $1$ & $1$ & $0$ & $-1$\\\hline
			&&&&&&\\[-4.5mm]
			\color{darkblue}$p_{\infty}^{+}=[\infty,+]$ & $1$ & $m_{-}$ & $1$ & $1$ & $2-m_{-}$ & $1-m_{-}$\\\hline
			&&&&&&\\[-4.5mm]
			\color{darkgreen}$p_{\infty}^{-}=[\infty,-]$ & $1$ & $m_{+}$ & $1$ & $1$ & $2-m_{+}$ & $1-m_{+}$\\\hline
			&&&&&&\\[-4.5mm]
			\color{darkblue}$g^{-1}(0)\setminus\{p_{\infty}^{+}\}$ & $\Omega-m_{+}$ & $1$ & $1$ & $-1$ & $-1$ & $0$\\\hline
			&&&&&&\\[-4.5mm]
			\color{darkgreen}$g^{-1}(\infty)\setminus\{p_{\infty}^{-}\}$ & $\Omega-m_{-}$ & $1$ & $1$ & $-1$ & $-1$ & $0$\\\hline
			&&&&&&\\[-4.5mm]
			\begin{tabular}{c}Trivial zeroes\\$g^{-1}(1)\setminus\{o\}$\end{tabular} & $\Omega-1$ & $1$ & $1$ & $0$ & $2$ & $2$\\\hline
			&&&&&&\\[-4.5mm]
			\begin{tabular}{c}Non-trivial zeroes\\$[\lambda_{*},+]$ or $[\lambda_{*},-]$\\$\lambda_{*}\neq0\in\Sp M_{\times}$\\$g_{*}$ as in (\ref{g* zero N})\end{tabular} & $\Omega-1$ & $1$ & $1$ & $0$ & $1$ & $1$\\\hline
			&&&&&&\\[-4.5mm]
			\begin{tabular}{c}Non-trivial zeroes\\$[\lambda_{*},+]$ or $[\lambda_{*},-]$\\$\lambda_{*}\neq0\in\Sp M_{\times}^{\rm R}$\\$g_{*}$ as in (\ref{gR* zero N})\end{tabular} & $\Omega-1$ & $1$ & $1$ & $0$ & $1$ & $1$\\\hline
		\end{tabular}
	\end{center}\vspace{-5mm}
	\caption{Poles and (trivial and non-trivial) zeroes of the meromorphic functions $\mathcal{N}_{\rm st}$ and $\mathcal{M}_{\rm st}$ defined in (\ref{N}), (\ref{M[N]}) for a \emph{non-reversible} counting process $Q_{t}$ satisfying assumption (A5) (and the additional generic assumption that the ramification points for $g$ have $e_{g}=2$, which is not used in the following). In the last three columns, the positive numbers are the orders of the zeroes, the negative numbers minus the orders of the poles, and $0$ indicates a point which is neither a zero nor a pole. Colours are as in figure~\ref{fig POI}.}
	\label{table poles zeroes Nst}
\end{table}

\begin{table}
	\begin{center}
		\begin{tabular}{|c|c||c|c||c|c|c|}\hline
			&&&&&&\\[-4.5mm]
			Point $p$ & Number & $e_{g}$ & $e_{\lambda}$ & $\frac{\dd\log g}{\dd\lambda}$ & $\mathcal{M}_{\rm st}$ & $\mathcal{N}_{\rm st}$\\\hline
			&&&&&&\\[-4.5mm]
			\color{darkred}$o=\overline{o}=p_{2\Omega}$ & $1$ & $1$ & $2$ & $-1$ & $-1$ & $0$\\\hline
			&&&&&&\\[-4.5mm]
			\begin{tabular}{c}Trivial zeroes\\\color{darkred}$p_{i}\in g^{-1}(1)\setminus\{o\}$\end{tabular} & $\Omega-1$ & $1$ & $2$ & $-1$ & $1$ & $2$\\\hline
			&&&&&&\\[-4.5mm]
			\color{darkred}$p_{i}\in g^{-1}(-1)$ & $\Omega$ & $1$ & $2$ & $-1$ & $-1$ & $0$\\\hline
			&&&&&&\\[-4.5mm]
			\begin{tabular}{c}Ramification\\points for $g$\\$[\lambda_{*},+]$ and $[\lambda_{*},-]$\\$\lambda_{*}\neq\infty$ as in (\ref{ramification points g})\end{tabular} & \begin{tabular}{l}$4\Omega-m_{+}$\\\;$-m_{-}-2$\end{tabular} & $2$ & $1$ & $1$ & $0$ & $-1$\\\hline
			&&&&&&\\[-4.5mm]
			\color{darkblue}$p_{\infty}^{+}=[\infty,+]$ & $1$ & $m_{-}$ & $1$ & $1$ & $2-m_{-}$ & $1-m_{-}$\\\hline
			&&&&&&\\[-4.5mm]
			\color{darkgreen}$p_{\infty}^{-}=[\infty,-]$ & $1$ & $m_{+}$ & $1$ & $1$ & $2-m_{+}$ & $1-m_{+}$\\\hline
			&&&&&&\\[-4.5mm]
			\color{darkblue}$g^{-1}(0)\setminus\{p_{\infty}^{+}\}$ & $\Omega-m_{+}$ & $1$ & $1$ & $-1$ & $-1$ & $0$\\\hline
			&&&&&&\\[-4.5mm]
			\color{darkgreen}$g^{-1}(\infty)\setminus\{p_{\infty}^{-}\}$ & $\Omega-m_{-}$ & $1$ & $1$ & $-1$ & $-1$ & $0$\\\hline
			&&&&&&\\[-4.5mm]
			\begin{tabular}{c}Non-trivial zeroes\\$[\lambda_{*},+]$ and $[\lambda_{*},-]$\\$\lambda_{*}\neq0\in\Sp M_{\times}$\end{tabular} & $2\Omega-2$ & $1$ & $1$ & $0$ & $1$ & $1$\\\hline
		\end{tabular}
	\end{center}\vspace{-5mm}
	\caption{Poles and (trivial and non-trivial) zeroes of the meromorphic functions $\mathcal{N}_{\rm st}$ and $\mathcal{M}_{\rm st}$ defined in (\ref{N}), (\ref{M[N]}) for a \emph{reversible} counting process $Q_{t}$ satisfying assumption (A5) (and the additional generic assumption that the ramification points for $g$ have $e_{g}=2$, which is not used in the following). The main modification compared to the non-reversible case of table~\ref{table poles zeroes Nst} is that the trivial zeroes are ramified for $\lambda$. In the last three columns, the positive numbers are the orders of the zeroes, the negative numbers minus the orders of the poles, and $0$ indicates a point which is neither a zero nor a pole. Colours are as in figure~\ref{fig POI}.}
	\label{table poles zeroes Nst reversible}
\end{table}

\subsection{Non-trivial zeroes of \texorpdfstring{$\mathcal{N}_{\rm st}$}{Nst}}
\label{section non-trivial zeroes Nst}
We consider the column vector $|U(g)\rangle$ and the row vector $\langle V(g)|$ defined as
\begin{eqnarray}
\label{UV}
&& |U(g)\rangle=|1\rangle\,g^{-1}-\frac{\sum_{k\in\Sout}|k\rangle\,w_{k\leftarrow 1}}{\wout}\nonumber\\
&& \langle V(g)|=g\,\langle1|-\frac{\sum_{j\in\Sin}w_{1\leftarrow j}\,\langle j|}{\wout}\;,
\end{eqnarray}
where $|\ell\rangle$, $\langle\ell|$ are the column and row vectors with $\ell$-th entry equal to $1$ and the others to $0$, and
\begin{equation}
\label{wout}
\wout=\sum_{k\in\Sout}w_{k\leftarrow1}\;.
\end{equation}
Then, we define the matrix
\begin{equation}
\label{Mtilde}
M_{\times}=M+\wout\,|U(1)\rangle\langle V(1)|\;.
\end{equation}
We observe that $M_{\times}$ is a Markov matrix. The corresponding Markov process, which we call the modified process (with respect to $Q_{t}$), does not have transitions $\Sin\to1$ or $1\to\Sout$. In particular, if $Q_{t}$ is the current between states $1$ and $j$, i.e. $\Sin=\Sout=\{j\}$, the modified process is simply the original process with the transitions $1\to j$ and $j\to1$ removed. Otherwise, assuming that all the rates $w_{k\leftarrow j}$ of the original process are non-zero, some transition rates are also changed. Incidentally, in the reversible case, detailed balance implies that the modified process has the same stationary state as the original process. This is not the case in general if $Q_{t}$ is not reversible.

\begin{lemma}
\label{lemma zeroes N 1}
Let $\lambda_{*}$ be a non-zero eigenvalue of $M_{\times}$ and $|\psi_{*}\rangle$ the corresponding right eigenvector of $M_{\times}$. Then, the unique point $p_{*}\in\R$ such that $\lambda(p_{*})=\lambda_{*}$ and $g(p_{*})=g_{*}$ with
\begin{equation}
\label{g* zero N}
g_{*}=\frac{\sum_{j\in\Sin}w_{1\leftarrow j}\,\langle j|\psi_{*}\rangle}{\wout\,\langle 1|\psi_{*}\rangle}
\end{equation}
is a zero of $\mathcal{N}$ for arbitrary initial condition $|P_{0}\rangle$ (under the assumption that $g_{*}$ is finite, non-zero and not a branch point for $g$).
\end{lemma}

\noindent\textbf{Remark:} the requirement that $\lambda(p_{*})$ is equal to an eigenvalue $\lambda_{*}\in\Sp M_{\times}$ specifies almost entirely the point $p_{*}\in\R$, as $p_{*}=[\lambda_{*},\pm]$. The identity (\ref{g* zero N}) for the value of $g(p_{*})$ simply amounts to choosing the correct sheet $\Ch_{+}$ or $\Ch_{-}$. Additionally, $g_{*}$ in (\ref{g* zero N}) can in principle be written as a rational function of $\lambda_{*}$ by solving the eigenvalue equation $M_{\times}|\psi_{*}\rangle=\lambda_{*}|\psi_{*}\rangle$ for the components of the eigenvector. Incidentally, if $\lambda_{*}=0$, one has necessarily $p_{*}=\overline{o}$ since $M_{\times}$ generically has a different stationary state as $M$.\\

\noindent
\textit{Proof of lemma \ref{lemma zeroes N 1}:} for any $g\in\C$, one has the identity
\begin{equation}
\label{M(g)[Mtilde]}
M(g)=M_{\times}-\wout\,|U(g)\rangle\langle V(g)|
\end{equation}
between the generator $M(g)$ of $Q_{t}$ and the fixed Markov matrix $M_{\times}$ independent of $g$ defined in (\ref{Mtilde}), with $U(g)$, $V(g)$ and $\wout$ defined in (\ref{UV}), (\ref{wout}). Since $|\psi_{*}\rangle$ is by definition an eigenvector of $M_{\times}$ with eigenvalue $\lambda_{*}$, (\ref{M(g)[Mtilde]}) implies $M(g)|\psi_{*}\rangle=\lambda_{*}|\psi_{*}\rangle-\big(g\,\wout\langle1|\psi_{*}\rangle-\sum_{j\in\Sin}w_{1\leftarrow j}\,\langle j|\psi_{*}\rangle\big)\,|U(g)\rangle$. Defining $g_{*}=\sum_{j\in\Sin}\frac{w_{1\leftarrow j}\,\langle j|\psi_{*}\rangle}{\wout\,\langle1|\psi_{*}\rangle}$, the vector $|\psi_{*}\rangle$ is then also an eigenvector of $M(g_{*})$ if $g_{*}$ is finite and non-zero (so that all matrix elements of $M(g_{*})$ are finite). The pair $(\lambda_{*},g_{*})$ corresponds to a unique point $p_{*}$ of the Riemann surface $\R$.

Additionally, since $M_{\times}$ is a Markov matrix, $\langle\Sigma|M_{\times}=0$, and one has from (\ref{M(g)[Mtilde]}) $\langle\Sigma|M(g)|\psi\rangle=-\langle\Sigma|U(g)\rangle\big(g\,\wout\langle1|\psi\rangle-\sum_{j\in\Sin}w_{1\leftarrow j}\,\langle j|\psi\rangle\big)$ for any $g$ and $|\psi\rangle$. From the definition of $g_{*}$ above, $\langle\Sigma|M(g)|\psi\rangle$ then vanishes when $g=g_{*}$ and $|\psi\rangle=|\psi_{*}\rangle$. Using the fact proved above that $|\psi_{*}\rangle$ is an eigenvector of $M(g_{*})$ with eigenvalue $\lambda_{*}$ and assuming $\lambda_{*}\neq0$, one has finally $\langle\Sigma|\psi_{*}\rangle=\frac{\langle\Sigma|M(g)|\psi_{*}\rangle}{\lambda_{*}}=0$. From (\ref{N}), the function $\mathcal{N}$ has a zero at $p_{*}$ if no cancellation occurs between the numerator and the denominator, i.e. if $g_{*}$ is not an exceptional point where several eigenvectors coincide and $g$ is ramified (see section~\ref{section ramification g}).
\\\rightline{$\square$}\\

We consider now a modified version of the reverse Markov process defined in (\ref{wR}), with Markov matrix
\begin{equation}
\label{MRtilde}
M_{\times}^{\rm R}=M^{\rm R}+\wRin\,|U^{\rm R}(1)\rangle\langle V^{\rm R}(1)|\;,
\end{equation}
build from the vectors
\begin{eqnarray}
\label{UVR}
&& |U^{\rm R}(g)\rangle=|1\rangle\,g-\frac{\sum_{j\in\Sin}|j\rangle\,w_{j\leftarrow 1}^{\rm R}}{\wRin}\nonumber\\
&& \langle V^{\rm R}(g)|=g^{-1}\,\langle1|-\frac{\sum_{k\in\Sout}w_{1\leftarrow k}^{\rm R}\,\langle k|}{\wRin}\;.
\end{eqnarray}
The rates $w_{k\leftarrow j}^{\rm R}$ are defined in (\ref{wR}) and
\begin{equation}
\label{wRin}
\wRin=\sum_{j\in\Sin}w_{j\leftarrow 1}^{\rm R}\;.
\end{equation}

\begin{lemma}
\label{lemma zeroes N 2}
Let $\lambda_{*}$ be a non-zero eigenvalue of $M_{\times}^{\rm R}$ and $|\psi_{*}^{\rm R}\rangle$ the corresponding right eigenvector of $M_{\times}^{\rm R}$. Then, the unique point $p_{*}\in\R$ such that $\lambda(p_{*})=\lambda_{*}$ and $g(p_{*})=g_{*}$ with
\begin{equation}
\label{gR* zero N}
g_{*}^{-1}=\frac{\sum_{k\in\Sout}w_{1\leftarrow k}^{\rm R}\,\langle k|\psi_{*}^{\rm R}\rangle}{\wRin\,\langle 1|\psi_{*}^{\rm R}\rangle}
\end{equation}
is a zero of $\mathcal{N}_{\rm st}$ (under the assumption that $g_{*}$ is finite, non-zero and not a branch point for $g$).
\end{lemma}

\noindent\textbf{Remark:} unlike in lemma~\ref{lemma zeroes N 1}, the eigenvalues of $M_{\times}^{\rm R}$ correspond to zeroes of $\mathcal{N}$ for \emph{stationary initial condition} only.\\

\noindent
\textit{Proof:} essentially the same as for lemma~\ref{lemma zeroes N 1}, based on the identity
\begin{equation}
\label{MR(g)[MRtilde]}
M^{\rm R}(g)=M_{\times}^{\rm R}-\wRin\,|U^{\rm R}(g)\rangle\langle V^{\rm R}(g)|
\end{equation}
with $M^{\rm R}(g)$ defined in (\ref{MR(g)}). As before, (\ref{MR(g)[MRtilde]}) implies that $|\psi_{*}^{\rm R}\rangle$ is also an eigenvector of $M^{\rm R}(g_{*})$ for $g_{*}$ defined through (\ref{gR* zero N}), which leads to $\langle\Sigma|M^{\rm R}(g_{*})|\psi_{*}^{\rm R}\rangle=0$ since $M_{\times}^{\rm R}$ is a Markov matrix and $\lambda_{*}\neq0$. Using the definition (\ref{MR(g)}) of $M^{\rm R}(g)$, transposition then gives $\langle\psi_{*}^{\rm R}|D_{\rm st}^{-1}M(g_{*})D_{\rm st}|\Sigma\rangle=0$, where $\langle\psi_{*}^{\rm R}|$ and $|\Sigma\rangle$ are the transpose of $|\psi_{*}^{\rm R}\rangle$ and $\langle\Sigma|$. Using $D_{\rm st}|\Sigma\rangle=|P_{\rm st}\rangle$ and calling $\langle\psi_{*}|=\langle\psi_{*}^{\rm R}|D_{\rm st}^{-1}$ leads to $\langle\psi_{*}|M(g_{*})|P_{\rm st}\rangle=0$. Finally, since $|\psi_{*}^{\rm R}\rangle$ is an eigenvector of $M(g_{*})$, (\ref{MR(g)}) implies that $\langle\psi_{*}|$ is an eigenvector of $M(g_{*})$, and the point $p_{*}$ is a zero of $\mathcal{N}_{\rm st}$, barring accidental cancellations when $g$ is ramified at $g_{*}$.
\\\rightline{$\square$}\\

Under the first part of assumption (A4), lemmas~\ref{lemma zeroes N 1} and \ref{lemma zeroes N 2} give $2\gen$ distinct points $p_{*}\in\R$, which are the non-trivial zeroes of $\mathcal{N}_{\rm st}$ if $g_{*}\neq1$. In the reversible case, we observe by comparison between (\ref{UV}) and (\ref{UVR}) that $|U^{\rm R}(g)\rangle=|U(g^{-1})\rangle$ and $\langle V^{\rm R}(g)|=\langle V(g^{-1})|$, which leads to $M_{\times}^{\rm R}=M_{\times}$. Then, taking the same eigenvalue $\lambda_{*}$ in lemmas~\ref{lemma zeroes N 1} and \ref{lemma zeroes N 2}, we observe that $g(p_{*})$ in (\ref{gR* zero N}) and (\ref{g* zero N}) are inverse of each other, and the corresponding points $p_{*}$ are thus related by the hyperelliptic involution (\ref{hyperelliptic involution}).

\subsection{Exact expression for \texorpdfstring{$\mathcal{M}_{\rm st}$}{Mst}, reversible case}
\label{section exact formula Mst reversible}
The poles (respectively zeroes) $p\in\R$ of a meromorphic function $f:\R\to\Ch$ with given orders $m(p)\geq1$ are the poles of the meromorphic differential $\dd\log f$, which are all simple, and have residues $-m(p)$ (resp. $m(p)$). In terms of formal linear combinations of points, this property can be conveniently written as
\begin{equation}
\label{Df=Rdlogf}
D(f)=R(\dd\log f)\;,
\end{equation}
where $D(f)$ (called the divisor of $f$) is the linear combination of the poles and zeroes of the function $f$ with coefficients equal to the orders of the zeroes and minus the orders of the poles, and $R(\omega)$ the linear combination of the poles of the meromorphic differential $\omega$ with coefficients equal to the corresponding residues.

Since a meromorphic differential of the form $\dd\log f$ is uniquely determined by the location of its poles and their residues, exhibiting $\omega$ with $R(\omega)=R(\dd\log\mathcal{M}_{\rm st})$, and such that $\exp(\int_{p_{0}}^{p}\omega)$ is a meromorphic function of $p$, implies $\mathcal{M}_{\rm st}(p)=\mathcal{M}_{\rm st}(p_{0})\,\exp(\int_{p_{0}}^{p}\omega)$. In the rest of this section, we apply this strategy in the reversible case, which is especially simple since the set of poles of $\dd\log\mathcal{M}_{\rm st}$ is symmetric under the hyperelliptic involution, unlike in the non-reversible case treated in section~\ref{section exact formula Mst}.

We consider the case of reversible $Q_{t}$, i.e. with $\Sin=\Sout$ and transition rates obeying $w_{k\leftarrow j}=w_{j\leftarrow k}^{\rm R}$. We additionally require that the generic assumption (A5) is satisfied, so that all the points in table~\ref{table poles zeroes Nst reversible} are distinct. Then, $m_{\pm}=3$, and gathering the results of the two previous sections (see in particular table~\ref{table poles zeroes Nst reversible}), we observe that $\mathcal{M}_{\rm st}$ only has simple poles, located at the $2\Omega$ ramification points $p_{i}$ for $\lambda$ and at the $2\Omega-4$ points in $g^{-1}(0)\cup g^{-1}(\infty)$ (including the two points $p_{\infty}^{\pm}$). On the other hand, $\mathcal{M}_{\rm st}$ has double zeroes at the points in $g^{-1}(1)\setminus\{o\}$ and simple zeroes at the points $[\lambda_{*},\pm]$ with $\lambda_{*}\in(\Sp M_{\times})\setminus\{0\}=(\Sp M_{\times}^{\rm R})\setminus\{0\}$. The divisor of $\mathcal{M}_{\rm st}$, see below (\ref{Df=Rdlogf}), is then equal to
\begin{eqnarray}
\label{divisor Mst reversible}
D\big(\mathcal{M}_{\rm st}\big)&=&
-\sum_{i=1}^{2\Omega}p_{i}-\sum_{p\in g^{-1}(0)}p-\sum_{p\in g^{-1}(\infty)}p\\
&&+2\sum_{p\in g^{-1}(1)\setminus\{o\}}p+\sum_{\lambda_{*}\in\Sp M_{\times}\setminus\{0\}}([\lambda_{*},+]+[\lambda_{*},-])\;.\nonumber
\end{eqnarray}
Note that the $\Omega-1$ points in $g^{-1}(1)\setminus\{o\}$ come twice in (\ref{divisor Mst reversible}) since they coincide with some of the $p_{i}$.

Since in the reversible case, the hyperelliptic involution $p\to\overline{p}$ sends $g(p)\to g(\overline{p})=1/g(p)$, the zeroes of $\mathcal{M}_{\rm st}$ in $g^{-1}(1)\setminus\{o\}$ come in pairs $[\lambda,+]$, $[\lambda,-]$. This is also the case for the poles in $g^{-1}(0)\cup g^{-1}(\infty)$, and the zeroes of the form $[\lambda_{*},\pm]$ with $\lambda_{*}\in(\Sp M_{\times})\setminus\{0\}$ since the values $g(p_{*})$ in lemmas~\ref{lemma zeroes N 1} and \ref{lemma zeroes N 2} are inverse of each other. The poles of $\dd\log\mathcal{M}_{\rm st}$ are then either the fixed points $p_{j}$ of the hyperelliptic involution or come by pairs with the same value of $\lambda$.

We note that for any finite $\lambda_{*}\in\C$, the poles of the meromorphic differential $\dd\lambda/(\lambda-\lambda_{*})$ are the two points $[\lambda_{*},\pm]$ if $\lambda_{*}$ is not ramified for $\lambda$ (respectively the single point $[\lambda_{*},+]=[\lambda_{*},-]=p_{i}$ for some $i$ if $\Delta(\lambda_{*})=0$), with residue $1$ (resp. residue $2$), and $p_{\infty}^{\pm}$, both with residue $-1$. Additionally, if $\lambda_{*}\neq0$, the function $p\mapsto\exp(\int_{o}^{p}\dd\lambda/(\lambda-\lambda_{*}))=1-\lambda(p)/\lambda_{*}$ is meromorphic on $\R$.

Similarly, for any $g_{*}\in\C$ not ramified for $g$, the poles of $\frac{\dd g}{g-g_{*}}$ are the points in $g^{-1}(g_{*})$, with residue $1$, the points in $g^{-1}(\infty)\setminus\{p_{\infty}^{-}\}$, non ramified for $g$, with residue $-1$, and the point $p_{\infty}^{-}$, ramified three times for $g$, with residue $-3$. For $g_{*}=0$, since $p_{\infty}^{+}\in g^{-1}(0)$ is ramified three times, its residue is equal to $3$ instead of $1$. If $g_{*}\neq1$, the function $p\mapsto\exp(\int_{o}^{p}\dd g/(g-g_{*}))=(g(p)-g_{*})/(1-g_{*})$ is again meromorphic on $\R$.

From the discussion above, one has
\begin{eqnarray}
\label{divisors reversible}
&&\hspace*{-5mm} R\Big(\sum_{\lambda_{*}\in\Sp M_{\times}\setminus\{0\}}\frac{\dd\lambda}{\lambda-\lambda_{*}}\Big)=-(\Omega-1)(p_{\infty}^{+}+p_{\infty}^{-})+\!\!\!\!\!\!\sum_{\lambda_{*}\in\Sp M_{\times}\setminus\{0\}}\!\!\!\!\!\!([\lambda_{*},+]+[\lambda_{*},-])\nonumber\\
&&\hspace*{-5mm} R\Big(-\frac{\dd\lambda}{\lambda}\Big)=-2o+p_{\infty}^{+}+p_{\infty}^{-}\nonumber\\
&&\hspace*{-5mm} R\Big(-\frac{\dd y}{y}\Big)=\Omega(p_{\infty}^{+}+p_{\infty}^{-})-\sum_{i=1}^{2\Omega}p_{i}\\
&&\hspace*{-5mm} R\Big(-\frac{\dd g}{g}\Big)=3p_{\infty}^{-}-3p_{\infty}^{+}+\sum_{p\in g^{-1}(\infty)\setminus\{p_{\infty}^{-}\}}p-\sum_{p\in g^{-1}(0)\setminus\{p_{\infty}^{+}\}}p\nonumber\\
&&\hspace*{-5mm} R\Big(\frac{2\,\dd g}{g-1}\Big)=-6p_{\infty}^{-}-2\sum_{p\in g^{-1}(\infty)\setminus\{p_{\infty}^{-}\}}p+2\sum_{p\in g^{-1}(1)}p\;,\nonumber
\end{eqnarray}
with the formal linear combination of poles $R(\omega)$ defined below (\ref{Df=Rdlogf}).

By inspection, we observe that the divisor (\ref{divisor Mst reversible}) is equal to the sum of the $R(\omega)$ in (\ref{divisors reversible}), and thus
\begin{equation}
\label{dlogMst reversible}
\dd\log\mathcal{M}_{\rm st}=\frac{2\,\dd g}{g-1}-\frac{\dd g}{g}-\frac{\dd y}{y}-\frac{\dd\lambda}{\lambda}
+\sum_{\lambda_{*}\in\Sp M_{\times}\setminus\{0\}}\frac{\dd\lambda}{\lambda-\lambda_{*}}\;.
\end{equation}
Using $\frac{\dd y}{y}+\frac{\dd g}{g}=\frac{\dd g}{g+1}+\frac{\dd g}{g-1}+\frac{P_{+}'(\lambda)}{P_{+}(\lambda)}\,\dd\lambda$, which follows from (\ref{y[g,lambda] sym}) with $P_{+}=P_{-}$, this can be rewritten as
\begin{equation}
\dd\log\mathcal{M}_{\rm st}=\frac{\dd g}{g-1}-\frac{\dd g}{g+1}+\Big(\tr\big((\lambda\Id-M_{\times})^{-1}\big)-\frac{2}{\lambda}-\frac{P_{+}'(\lambda)}{P_{+}(\lambda)}\Big)\dd\lambda\;.
\end{equation}
After integration and matching the behaviour $p\to o$, one finds the following identity.

\begin{lemma}
In the reversible case with stationary initial condition, one has under assumption (A5)
\begin{equation}
\label{Mst reversible}
\mathcal{M}_{\rm st}(p)=\frac{1-g}{1+g}\,\frac{P_{+}(0)}{P_{+}(\lambda)}\,\frac{(-1)^{\Omega}\det(\lambda\Id-M_{\times})}{\lambda^{2}\prod_{\lambda_{*}\in\Sp M_{\times}\setminus\{0\}}\lambda_{*}}\;,
\end{equation}
where $\lambda$ and $g$ are evaluated at the point $p$.
\end{lemma}

\noindent\textbf{Remark:} at this point, assumption (A5) can largely be relaxed by continuity of $\mathcal{M}_{\rm st}([\lambda,\pm])$ with respect to the transition rates $w_{k\leftarrow j}$, assuming $P_{+}=P_{-}$ does not vanish identically, which is necessary for the Riemann surface $\R$ to be defined at all. When the eigenvalue $0$ of $M_{\times}$ is degenerate, (\ref{Mst reversible}) must however be computed by taking properly the limit from a non-degenerate case.

\noindent\textbf{Remark:} the identity $\frac{1-g}{1+g}=\frac{y}{P_{0}(\lambda)-2P_{+}(\lambda)}$ leads to the canonical algebraic form (\ref{canonical form mero function}) with $R=0$ for $\mathcal{M}_{\rm st}$.

The expansion of $\mathcal{M}_{\rm st}$ (or rather $\mathcal{N}_{\rm st}$) near the point $o$ gives access to the late time correction to stationary large deviations. In particular, we obtain
\begin{equation}
\label{Var(Qt) late times}
\langle Q_{t}^{2}\rangle\simeq D(t+t_{\times}^{\rm K})+\frac{\lambda_{\rm st}^{(4)}(0)}{6D}-\frac{1}{6}-D\,\frac{P_{+}'(0)}{P_{+}(0)}\;,
\end{equation}
up to exponentially small corrections at late times. The constant term depends on the fourth stationary cumulant $\lambda_{\rm st}^{(4)}(0)$ of $Q_{t}$, and also on Kemeny's constant for the modified process $t_{\times}^{\rm K}=-\sum_{\lambda_{*}\in\Sp M_{\times}\setminus\{0\}}\frac{1}{\lambda_{*}}>0$, which is also the expectation value for any initial state $j$ of the first passage time from $j$ to $k$ averaged on $k$ with the stationary measure, see e.g. \cite{BHLMT2018.1}. We emphasize that (\ref{Var(Qt) late times}) is only valid for a system prepared initially in the stationary initial condition: the correction to stationary large deviations depends in general on the initial condition.

The time-dependent probability of $Q_{t}$ is given by (\ref{P[int R]}). In the reversible case, with stationary initial condition, (\ref{Mst reversible}) leads to
\begin{equation}
\label{P[int R] reversible}
\P(Q_{t}=Q)=\frac{1}{2\ii\pi}\oint_{\Gamma}\dd\lambda\,\frac{\ee^{t\lambda}}{g(\lambda)^{Q}}\,\frac{1-g(\lambda)}{1+g(\lambda)}\,\frac{P_{+}(0)}{P_{+}(\lambda)}\,\frac{(-1)^{\Omega}\det(\lambda\Id-M_{\times})}{\lambda^{2}\prod_{\lambda_{*}\in\Sp M_{\times}\setminus\{0\}}\lambda_{*}}\;,
\end{equation}
where $g(\lambda)=\frac{y(\lambda)-P_{0}(\lambda)}{2P_{+}(\lambda)}$, $y(\lambda)=\prod_{i=1}^{\Omega}\big((\lambda-\lambda_{2i-1})\sqrt{\frac{\lambda-\lambda_{2i}}{\lambda-\lambda_{2i-1}}}\big)$ with the usual determination of the square root, and $\Gamma$ is a closed curve with negative orientation enclosing the cuts $(\lambda_{2i-1},\lambda_{2i})$, $i=1,\ldots,\Omega$ and excluding the zeroes of $P_{+}$. The contour integral in (\ref{P[int R] reversible}) can also be computed from the residues of the essential singularity at infinity and of the poles at the zeroes of $P_{+}$.

Alternatively, observing that $y(\lambda)\to\pm\ii(-1)^{\Omega-\ell}s_{\ell}(\lambda)$ when $\lambda$ converges to the cut $(\lambda_{2\ell-1},\lambda_{2\ell})$ from above ($\Im\lambda>0$; sign $+$) or below ($\Im\lambda<0$; sign $-$), with
\begin{equation}
\label{sl(lambda)}
s_{\ell}(\lambda)=\prod_{j=1}^{2\ell-1}\sqrt{\lambda-\lambda_{j}}\prod_{j=2\ell}^{2\Omega}\sqrt{\lambda_{j}-\lambda}\;,
\end{equation}
the contour integral around each cut in (\ref{P[int R] reversible}) can be written as the integral of $2\ii$ times the imaginary part of the integrand on the cut (reached from above). Using $\frac{1-g}{1+g}=\frac{y}{P_{0}(\lambda)-2P_{+}(\lambda)}$, this leads after some rewriting to the following expression in terms of real integrals on the cuts.

\begin{prop}
\label{prop proba Qt reversible}
In the reversible case, the time-dependent probability of $Q_{t}$ with stationary initial condition is given by
\begin{eqnarray}
\label{P[int cut] reversible}
&&\hspace*{-9mm}\P(Q_{t}=Q)=\sum_{\ell=1}^{\Omega}\frac{(-1)^{\ell}}{\pi}\int_{\lambda_{2\ell-1}}^{\lambda_{2\ell}}\!\!\dd\lambda\,\frac{s_{\ell}(\lambda)\,P_{+}(0)/P_{+}(\lambda)}{P_{0}(\lambda)-2P_{+}(\lambda)}\,\frac{\ee^{t\lambda}\,\det(\lambda\Id-M_{\times})}{\lambda^{2}\prod_{\lambda_{*}\in\Sp M_{\times}\setminus\{0\}}\lambda_{*}}\\
&&\hspace*{75mm}\times\Re\Big[\Big(\frac{2P_{+}(\lambda)}{\ii s_{\ell}(\lambda)-P_{0}(\lambda)}\Big)^{Q}\Big]\;,\nonumber
\end{eqnarray}
with $P_{0}$ and $P_{+}=P_{-}$ the polynomials defined from (\ref{det}), $\lambda_{1}<\ldots<\lambda_{2\Omega}$ the zeroes of the polynomial $\Delta(\lambda)=P_{0}(\lambda)^{2}-4P_{+}(\lambda)^{2}$, $M_{\times}$ the matrix defined in (\ref{Mtilde}) and $s_{\ell}(\lambda)$ given by (\ref{sl(lambda)}). The identity (\ref{P[int cut] reversible}) requires the generic assumptions that the $\lambda_{i}$ are distinct (and that the eigenvalue $0$ of $M_{\times}$ is not degenerate; otherwise the integrand has to be computed as the limit from a non-degenerate case).
\end{prop}

\noindent\textbf{Remark:} the integral in (\ref{P[int cut] reversible}) is well defined if the $\lambda_{i}$ are distinct. Indeed, since $(P_{0}(\lambda)-2P_{+}(\lambda))(P_{0}(\lambda)+2P_{+}(\lambda))=\Delta(\lambda)$, the poles coming from the denominator $P_{0}(\lambda)-2P_{+}(\lambda)$ partially cancel with factors in $s_{\ell}(\lambda)$, giving integrable inverse square root singularities. Similarly, the denominator $\lambda^{2}$ partially cancels with a factor $\lambda$ in $\det(\lambda\Id-M_{\times})$ (since $M_{\times}$ is a Markov matrix) and a factor $\sqrt{-\lambda}$ in $s_{\ell}(\lambda)$ since $\lambda_{2\Omega}=0$. Finally, $(\ii s_{\ell}(\lambda)-P_{0}(\lambda))(-\ii s_{\ell}(\lambda)-P_{0}(\lambda))=4P_{+}(\lambda)^{2}$ and $P_{+}(\lambda)$ can not vanish on the cuts since the points on the cuts have $|g|=1$ while the zeroes of $P_{+}$ have $g=\infty$.

\noindent\textbf{Remark:} in the reversible case with stationary initial condition, one has $\P(Q_{t}=Q)=\P(Q_{t}=-Q)$ at any time $t$. From (\ref{P[int R]}), it follows by making the change of variable $h\to h^{-1}$ and using $M(h^{-1})=M^{\rm R}(h)$ and (\ref{MR(g)}). From (\ref{P[int cut] reversible}), it is a consequence of the identity $s_{\ell}(\lambda)^{2}=-\Delta(\lambda)$.

\subsection{Explicit reversible example with 3 states}
\label{section 3 states model}
\begin{figure}
	\begin{tabular}{lll}
		\begin{picture}(50,44)(0,-3)
			\put(25,40){\circle{7.5}}\put(24.1,38.75){$1$}
			\put(5,4){\circle{7.5}}\put(4.1,2.75){$2$}
			\put(45,4){\circle{7.5}}\put(44.1,2.75){$3$}
			\put(21,36){\thicklines\color{darkred}\vector(-11,-20){15}}\put(9,8){\thicklines\color{darkred}\vector(11,20){15}}\put(10,22){$1$}\put(17.5,18.5){$1$}
			\put(10,2.5){\vector(1,0){30}}\put(40,5.5){\vector(-1,0){30}}\put(25,-0.5){$q$}\put(25,7.5){$p$}
			\put(44.3,9){\vector(-11,20){15}}\put(26.5,35){\vector(11,-20){15}}\put(31,18.5){$q$}\put(39.25,22.5){$p$}
		\end{picture}
		&\hspace*{10mm}&
		\includegraphics[width=50mm]{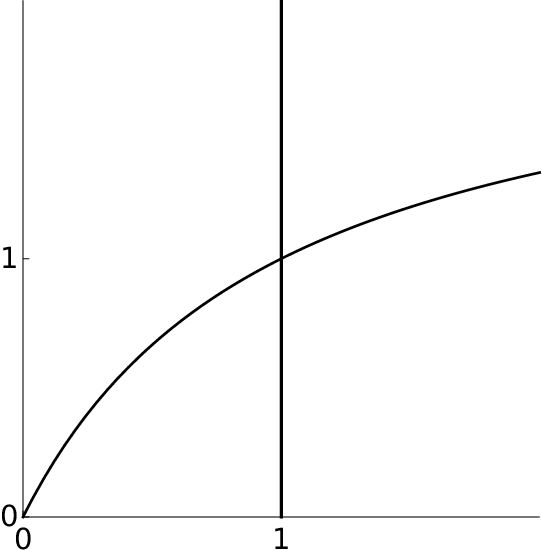}
		\begin{picture}(0,0)
			\put(-4,5){$p$}
			\put(-48,48){$q$}
			\put(-12,29){$q=\frac{2p}{p+1}$}
			\put(-35,10){I}\put(-15,15){II}\put(-40,35){III}\put(-15,40){IV}
		\end{picture}
	\end{tabular}
	\caption{Dynamics of the reversible three states model studied in section~\ref{section 3 states model} (left), and corresponding allowed sectors I, II, III, IV satisfying assumption (A2) for the transition rates $p$ and $q$ (right).}
	\label{fig 3 states model}
\end{figure}
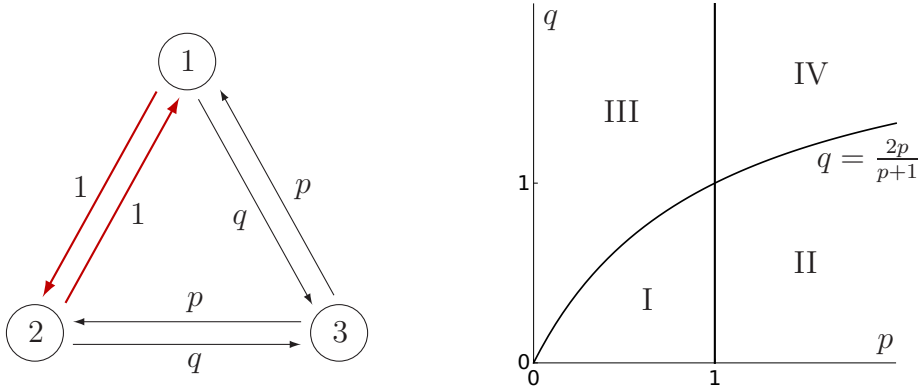

For concreteness, we consider in this section a reversible case with $\Omega=3$ states, and write explicitly the expression (\ref{P[int cut] reversible}) for the probability of the current $Q_{t}$ between states $1$ and $2$ (i.e. with $\Sin=\Sout=\{2\}$) in that case. Reversibility is then equivalent to $w_{2\leftarrow1}w_{3\leftarrow2}w_{1\leftarrow3}=w_{1\leftarrow2}w_{2\leftarrow3}w_{3\leftarrow1}$, but we restrict for simplicity to transition rates $w_{2\leftarrow1}=w_{1\leftarrow2}=1$, $w_{1\leftarrow3}=w_{2\leftarrow3}=p$, $w_{3\leftarrow1}=w_{3\leftarrow2}=q$, $p,q\geq0$, with generator
\begin{equation}
M(g)=\left(\begin{array}{ccc}
-1-q & g^{-1} & p\\
g & -1-q & p\\
q & q & -2p
\end{array}\right)\;,
\end{equation}
see figure~\ref{fig 3 states model}. The stationary state is then given by $P_{\rm st}(1)=P_{\rm st}(2)=\frac{p}{2p+q}$ and $P_{\rm st}(3)=\frac{q}{2p+q}$.

\begin{table}
	\begin{center}
		\begin{tabular}{c|cccccc}
			Sector & $\lambda_{1}$ & $\lambda_{2}$ & $\lambda_{3}$ & $\lambda_{4}$ & $\lambda_{5}$ & $\lambda_{6}$\\\hline
			I & $\lambda_{-}$ & $-2-q$ & $-2p-q$ & $\lambda_{+}$ & $-q$ & 0\\
			II & $\lambda_{-}$ & $-2p-q$ & $-2-q$ & $\lambda_{+}$ & $-q$ & 0\\
			III & $\lambda_{-}$ & $-2-q$ & $-2p-q$ & $-q$ & $\lambda_{+}$ & 0\\
			IV & $\lambda_{-}$ & $-2p-q$ & $-2-q$ & $-q$ & $\lambda_{+}$ & 0
		\end{tabular}
	\end{center}
	\caption{Properly ordered branch points $\lambda_{i}$ for the reversible three states model studied in section~\ref{section 3 states model}, depending on the sector in parameter space as in figure~\ref{fig 3 states model}. The branch points $\lambda_{\pm}=\frac{-2-2p-q\pm\sqrt{(2+2p+q)^{2}-16p}}{2}$ are the zeroes of the quadratic factor of the polynomial $\Delta$. The sectors are separated by the lines $p=1$ and $q=\frac{2p}{p+1}$ where $-2-q=-2p-q$ and $\lambda_{+}=-q$ respectively. The branch points $0$, $-2-q$ and $-2p-q$ (respectively $-q$, $\lambda_{+}$ and $\lambda_{-}$) are the eigenvalues of $M(1)$ (resp. $M(-1)$). Transitions between the sectors correspond to eigenvalue crossings for $M(1)$ or $M(-1)$.}
	\label{table branch points 3 states model}
\end{table}

For this simple case, one finds from (\ref{det}) $P_{+}(\lambda)=P_{-}(\lambda)=-pq$ and
\begin{equation}
P_{0}(\lambda)=\lambda^{3}+2(p+q+1)\lambda^{2}+(2+q)(2p+q)\lambda+2pq\;,
\end{equation}
which leads to $\Delta(\lambda)=\lambda(\lambda+q)(\lambda+2+q)(\lambda+2p+q)(\lambda^{2}+(2+2p+q)\lambda+4p)$. Assumption (A2) that the spectral curve (\ref{spectral curve}) is non-degenerate, i.e. that all the zeroes of $\Delta$ are distinct, is then equivalent to $p,q\neq0$, $p\neq1$, $q\neq\frac{2p}{p+1}$, which leads to the existence of four sectors, see figure~\ref{fig 3 states model}. The definitions of the roots $\lambda_{i}$ of $\Delta$ ordered as $\lambda_{1}<\ldots<\lambda_{6}$ depend on the sector, as shown in table~\ref{table branch points 3 states model}. As expected (see the paragraph of section~\ref{section ramification lambda} about the reversible case), each pair $\{\lambda_{2i-1},\lambda_{2i}\}$ is made of one eigenvalue of $M(1)$ and one eigenvalue of $M(-1)$.

Using $\det(\lambda\,\Id-M_{\times})=\lambda(\lambda+q)(\lambda+2p+q)$, the expression (\ref{P[int cut] reversible}) for the probability of $Q_{t}$ finally rewrites
\begin{eqnarray}
&&\hspace*{-10mm}\P(Q_{t}=Q)=\sum_{\ell=1}^{3}\frac{(-1)^{\ell}}{\pi}\int_{\lambda_{2\ell-1}}^{\lambda_{2\ell}}\!\!\!\dd\lambda\,\frac{(\lambda+2p+q)\,s_{\ell}(\lambda)\,\ee^{t\lambda}}{q(2p+q)\lambda(\lambda^{2}+(2+2p+q)\lambda+4p)}\\
&&\hspace{70mm}\times\Re\Big[\Big(\frac{2pq}{P_{0}(\lambda)-\ii s_{\ell}(\lambda)}\Big)^{Q}\Big]\;,\nonumber
\end{eqnarray}
where the definition of $s_{\ell}(\lambda)$, given in (\ref{sl(lambda)}), depends on the sector.

\subsection{Exact expression for \texorpdfstring{$\mathcal{M}_{\rm st}$}{Mst}, general case}
\label{section exact formula Mst}
In the general case, the poles and zeroes of $\mathcal{M}_{\rm st}$ no longer come by pairs $[\lambda_{*},\pm]$, and one has the divisor
\begin{eqnarray}
\label{divisor Mst}
&&\hspace*{-5mm}D\big(\mathcal{M}_{\rm st}\big)=
2\sum_{p\in g^{-1}(1)\setminus\{o\}}p
+\sum_{\substack{p\in\R,\\\lambda(p)=\lambda_{*},\;g(p)=g_{*}\\\lambda_{*}\in\Sp M_{\times}\setminus\{0\}\\g_{*}\text{ as in (\ref{g* zero N})}}}p
+\sum_{\substack{p\in\R,\\\lambda(p)=\lambda_{*},\;g(p)=g_{*}\\\lambda_{*}\in\Sp M_{\times}^{\rm R}\setminus\{0\}\\g_{*}\text{ as in (\ref{gR* zero N})}}}p\\
&&\hspace*{15mm}-\sum_{i=1}^{2\Omega}p_{i}
-\sum_{p\in g^{-1}(0)}p
-\sum_{p\in g^{-1}(\infty)}p
+(3-m_{-})p_{\infty}^{+}+(3-m_{+})p_{\infty}^{-}\;,\nonumber
\end{eqnarray}
see table~\ref{table poles zeroes Nst}.

Meromorphic differentials with prescribed simple poles on hyperelliptic Riemann surfaces can be constructed quite explicitly. Indeed, for any $\lambda_{*}\in\C$ not a ramification point for $\lambda$ (i.e. with $\Delta(\lambda_{*})\neq0$) and $y_{*}=\pm\sqrt{\Delta(\lambda_{*})}$, the poles of the meromorphic differential $\frac{y+y_{*}}{2y}\,\frac{\dd\lambda}{\lambda-\lambda_{*}}$, which are all simple, are $p_{*}=[\lambda_{*},\pm]$ (with the same sign as for $y_{*}$), such that $\lambda(p_{*})=\lambda_{*}$ and $y(p_{*})=y_{*}$, which has residue $1$, and both $p_{\infty}^{+}$ and $p_{\infty}^{-}$, with residue $-1/2$. Additionally, from (\ref{g(p)}), if $g(p_{*})=g_{*}$ is known, the differential above can be written as
\begin{equation}
\label{omega lambda* g*}
\omega_{\lambda_{*},g_{*}}(p)=\frac{y(p)+g_{*}\,P_{+}(\lambda_{*})-g_{*}^{-1}P_{-}(\lambda_{*})}{2y(p)}\,\frac{\dd\lambda(p)}{\lambda(p)-\lambda_{*}}\;.
\end{equation}
This has the advantage of being independent of the sign in the definition of $y_{*}$, which depends on the choice of the branch cuts $\gamma_{i}$.

Since the difference of two differentials of the form $\omega_{\lambda_{*},g_{*}}$ has two simple poles on $\R$, with residues $1$ and $-1$, located at arbitrary points of $\R$ with finite, un-ramified $\lambda$, they can be used as building block for some meromorphic differentials with simple poles. One has in particular
\begin{eqnarray}
\label{divisors}
&& R\Big(\sum_{\substack{\lambda_{*}\in\Sp M_{\times}\setminus\{0\}\\g_{*}\text{ as in (\ref{g* zero N})}}}\omega_{\lambda_{*},g_{*}}\Big)=-\frac{\Omega-1}{2}\,(p_{\infty}^{+}+p_{\infty}^{-})+\sum_{\substack{p\in\R,\\\lambda(p)=\lambda_{*},\;g(p)=g_{*}\\\lambda_{*}\in\Sp M_{\times}\setminus\{0\}\\g_{*}\text{ as in (\ref{g* zero N})}}}p\nonumber\\
&& R\Big(\sum_{\substack{\lambda_{*}\in\Sp M_{\times}^{\rm R}\setminus\{0\}\\g_{*}\text{ as in (\ref{gR* zero N})}}}\omega_{\lambda_{*},g_{*}}\Big)=-\frac{\Omega-1}{2}\,(p_{\infty}^{+}+p_{\infty}^{-})+\sum_{\substack{p\in\R,\\\lambda(p)=\lambda_{*},\;g(p)=g_{*}\\\lambda_{*}\in\Sp M_{\times}^{\rm R}\setminus\{0\}\\g_{*}\text{ as in (\ref{gR* zero N})}}}p\nonumber\\
&& R\Big(-2\,\omega_{0,1}\Big)=-2o+p_{\infty}^{+}+p_{\infty}^{-}\\
&& R\Big(-\frac{\dd y}{y}\Big)=\Omega(p_{\infty}^{+}+p_{\infty}^{-})-\sum_{i=1}^{2\Omega}p_{i}\nonumber\\
&& R\Big(-\frac{\dd g}{g}\Big)=m_{+}p_{\infty}^{-}-m_{-}p_{\infty}^{+}+\sum_{p\in g^{-1}(\infty)\setminus\{p_{\infty}^{-}\}}p-\sum_{p\in g^{-1}(0)\setminus\{p_{\infty}^{+}\}}p\nonumber\\
&& R\Big(\frac{2\,\dd g}{g-1}\Big)=-2m_{+}p_{\infty}^{-}-2\sum_{p\in g^{-1}(\infty)\setminus\{p_{\infty}^{-}\}}p+2\sum_{p\in g^{-1}(1)}p\;.\nonumber
\end{eqnarray}
We observe that the sum of the $R(\omega)$ in (\ref{divisors}) is equal to the divisor (\ref{divisor Mst}). A crucial difference with the reversible case is that while the meromorphic differentials in (\ref{divisors reversible}) are already of the form $\dd\log f$, this is not the case for the differentials $\omega_{\lambda_{*},g_{*}}$ in (\ref{divisors}). Thus, one can only deduce from (\ref{Df=Rdlogf}) that $\dd\log\mathcal{M}_{\rm st}$ minus the sum of the differentials in (\ref{divisors}) has no poles, and is thus a \emph{holomorphic} differential. Holomorphic differentials on a compact Riemann surface of genus $\gen$ form a $\gen$-dimensional vector space, generated in the hyperelliptic case considered here by the $\frac{\lambda^{\ell-1}\,\dd\lambda}{y}$, $\ell=1,\ldots,\gen$. The considerations above then lead to the following identity.

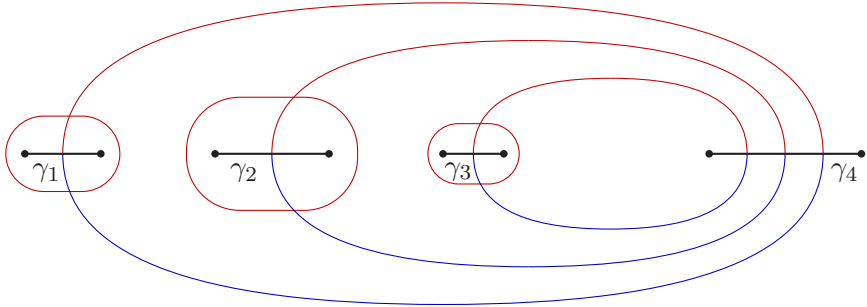
\begin{figure}
	\begin{center}
		\begin{picture}(80,40)
			{\thicklines
				\put(-15,20){\line(1,0){10}}\put(-15,20){\circle*{1}}\put(-5,20){\circle*{1}}\put(-14,17){\small$\gamma_{1}$}
				\put(10,20){\line(1,0){15}}\put(10,20){\circle*{1}}\put(25,20){\circle*{1}}\put(12,17){\small$\gamma_{2}$}
				\put(40,20){\line(1,0){8}}\put(40,20){\circle*{1}}\put(48,20){\circle*{1}}\put(40.2,17.2){\small$\gamma_{3}$}
				\put(75,20){\line(1,0){20}}\put(75,20){\circle*{1}}\put(95,20){\circle*{1}}\put(91,17){\small$\gamma_{4}$}
			}%
			{\color{darkred}
				\put(-10,20){\oval(15,10)}
				\put(17.5,20){\oval(22.5,15)}
				\put(44,20){\oval(12,8)}
				\qbezier(80,20)(80,30)(62,30)\qbezier(62,30)(44,30)(44,20)
				\qbezier(85,20)(85,35)(51.25,35)\qbezier(51.25,35)(17.5,35)(17.5,20)
				\qbezier(90,20)(90,40)(40,40)\qbezier(40,40)(-10,40)(-10,20)
			}%
			{\color{darkblue}
				\qbezier(80,20)(80,10)(62,10)\qbezier(62,10)(44,10)(44,20)
				\qbezier(85,20)(85,5)(51.25,5)\qbezier(51.25,5)(17.5,5)(17.5,20)
				\qbezier(90,20)(90,0)(40,0)\qbezier(40,0)(-10,0)(-10,20)
			}%
		\end{picture}
	\end{center}\vspace{-5mm}
	\caption{Standard basis of $2\gen$ loops for a hyperelliptic Riemann surface (plotted here with genus $\gen=3$) under some choice of branch cuts $\gamma_{i}$. The red and blue portions of the loops belong respectively to the sheet $\C_{+}$ and $\C_{-}$ of $\R$.}
	\label{fig basis loops}
\end{figure}

\begin{prop}
\label{prop proba Qt}
under generic assumption (A5), the differential of $\log\mathcal{M}_{\rm st}$ is given by
\begin{eqnarray}
\label{dlogMst}
&&
\dd\log\mathcal{M}_{\rm st}
=\sum_{\ell=1}^{\Omega-1}c_{\ell}\,\frac{\lambda^{\ell-1}\,\dd\lambda}{y}
+\frac{2\,\dd g}{g-1}-\frac{\dd g}{g}-\frac{\dd y}{y}
-2\,\omega_{0,1}\nonumber\\
&&\hspace*{27mm}
+\sum_{\substack{\lambda_{*}\in\Sp M_{\times}\setminus\{0\}\\g_{*}\text{ as in (\ref{g* zero N})}}}\omega_{\lambda_{*},g_{*}}
+\sum_{\substack{\lambda_{*}\in\Sp M_{\times}^{\rm R}\setminus\{0\}\\g_{*}\text{ as in (\ref{gR* zero N})}}}\omega_{\lambda_{*},g_{*}}\;,
\end{eqnarray}
The differential $\omega_{\lambda_{*},g_{*}}$ is defined for arbitrary $\lambda_{*},g_{*}$ in (\ref{omega lambda* g*}). The constants $c_{\ell}$ are not determined, see however below for a procedure allowing to fix them, which implies incidentally that all $c_{\ell}\in\mathbb{R}$, and for a conjecture about their sign.
\end{prop}

In the reversible case, $P_{+}=P_{-}$, $\Sp M_{\times}^{\rm R}=\Sp M_{\times}$ and the fact that $g_{*}$ in (\ref{g* zero N}) and (\ref{gR* zero N}) are inverse of each other lead to cancellations in (\ref{dlogMst}), and (\ref{dlogMst reversible}) is recovered by taking all $c_{\ell}=0$. In the non-reversible case, using (\ref{N(o)}), the constants $c_{\ell}$ ensure (assuming $J\neq0$, otherwise another base point as $o$ must be chosen for the integral in the exponential) that
\begin{equation}
\label{Mst[dlogMst]}
\mathcal{M}_{\rm st}(p)=J^{-1}\,\ee^{\int_{o}^{p}\dd\log\mathcal{M}_{\rm st}}
\end{equation}
is meromorphic on $\R$, and in particular that the non-analyticities of the factors $\sqrt{\det(\lambda\Id-M_{\times})}$ and $\sqrt{\det(\lambda\Id-M_{\times}^{R})}$ coming from the $y$-independent part of the $\omega_{\lambda_{*},g_{*}}$ disappear. The probability of $Q_{t}$ can then be computed, at least in principle, as (\ref{P[int R]}) with $\mathcal{M}_{\rm st}$ given by (\ref{Mst[dlogMst]}), (\ref{dlogMst}).

From (\ref{Mst[dlogMst]}), the function $\mathcal{M}_{\rm st}$ is meromorphic on $\R$ if and only if all the periods of $\dd\log\mathcal{M}_{\rm st}$, i.e. its integrals over closed curve on $\R$, are integer multiples of $2\ii\pi$. Since closed curves on $\R$ are generated by a basis of $2\gen$ independent simple closed curves \footnote{Modulo homology. Any simple closed curve that is a boundary, i.e. such that cutting $\R$ along it splits $\R$ into disjoint connected components, is a null curve, along which integrals of meromorphic differentials can be computed by residues.}, see figure \ref{fig basis loops} for the standard construction in the hyperelliptic case, the real and imaginary parts of the $\gen$ constants $c_{\ell}$ can be fixed so that the integrals over the $2\gen$ elements of the basis are purely imaginary. That the $c_{\ell}$ are additionally integer multiples of $2\ii\pi$ then follows automatically from the fact that (\ref{dlogMst}) is indeed equal to $\dd\log\mathcal{M}_{\rm st}$ by design.

Unlike in the reversible case, writing $\mathcal{M}_{\rm st}$ explicitly in canonical algebraic form (\ref{canonical form mero function}) from (\ref{Mst[dlogMst]}), (\ref{dlogMst}) is in general a difficult problem. Additionally, the function $\mathcal{M}_{\rm st}$ can in principle be written as a product of theta functions, which are infinite sums built from the period matrix of holomorphic differentials normalized with respect to a basis of closed curves, see e.g. \cite{K1999.2}.

The identity (\ref{d1 st}) implies $\frac{\dd\log\mathcal{N}_{\rm st}}{\dd\log g}(o)=0$, and (\ref{dlogMst}) then leads to the exact expression
\begin{equation}
\label{c1}
c_{1}=-\sum_{\substack{\lambda_{*}\in\Sp M_{\times}\setminus\{0\}\\g_{*}\text{ as in (\ref{g* zero N})}}}\frac{y\,\omega_{\lambda_{*},g_{*}}}{\dd\lambda}(o)
-\sum_{\substack{\lambda_{*}\in\Sp M_{\times}^{\rm R}\setminus\{0\}\\g_{*}\text{ as in (\ref{gR* zero N})}}}\frac{y\,\omega_{\lambda_{*},g_{*}}}{\dd\lambda}(o)\;,
\end{equation}
where $-\frac{y\,\omega_{\lambda_{*},g_{*}}}{\dd\lambda}(o)=\frac{P_{+}(0)-P_{-}(0)+g_{*}P_{+}(\lambda_{*})-g_{*}^{-1}P_{-}(\lambda_{*})}{2\lambda_{*}}$ explicitly. Additionally, the coefficient $c_{1}$ can in principle be written as a rational function of the transition rates $w_{k\leftarrow j}$ with coefficients in $\mathbb{Q}$. Indeed, solving the eigenvalue equation for the eigenvector $|\psi_{*}\rangle$ appearing in (\ref{g* zero N}) leads to a rational expression in the $w_{k\leftarrow j}$ and in $\lambda_{*}$ for its entries, and hence also for $g_{*}$ from (\ref{g* zero N}). Then $y_{*}=g_{*}P_{+}(\lambda_{*})-g_{*}^{-1}P_{-}(\lambda_{*})$ can be written as $y_{*}=Y(\lambda_{*})$ with $Y$ a rational function with coefficients rational in the $w_{k\leftarrow j}$, and
\begin{equation}
\sum_{\substack{\lambda_{*}\in\Sp M_{\times}\\g_{*}\text{ as in (\ref{g* zero N})}}}\frac{y\,\omega_{\lambda_{*},g_{*}}}{\dd\lambda}(p)=\frac{1}{2}\tr\Big(\frac{y(p)\Id+Y(M_{\times})}{\lambda(p)\Id-M_{\times}}\Big)\;.
\end{equation}
The same reasoning also applies to $g_{*}$ as in (\ref{gR* zero N}). From (\ref{c1}), this finally leads in the limit $p\to o$ to a rational expression in the $w_{k\leftarrow j}$ for $c_{1}$.

The argument above implies in particular that $c_{1}\in\mathbb{R}$. Generalization to any $c_{\ell}$ is straightforward by using (\ref{dlogMst}) and noting that the expansion near $p=o$ in the variable $\nu=\log g$ of $\lambda$ and of the corresponding eigenvectors of $M(g)$ (and hence also the expansion of $\mathcal{N}_{\rm st}$) have coefficients rational in the transition rates. Additionally, we conjecture from numerics that all the $c_{\ell}$ have the same sign as the late time average $J$ of $Q_{t}/t$.


\subsection{Almost homogeneous, symmetric 1d random walk}
\label{section simple random walk}
In order to illustrate more concretely the results of the previous section, we consider in this section the simple example of the one-dimensional random walk with transition rates
\begin{eqnarray}
&& w_{2\leftarrow1}=1+q\nonumber\\
&& w_{1\leftarrow2}=1-q\\
&& w_{j+1\leftarrow j}=w_{j\leftarrow j+1}=1\nonumber
\text{ for $j\neq1$}\;,
\end{eqnarray}
where $-1<q<1$, and define as before $Q_{t}$ as the current between states $1$ and $2$. For this simple non-reversible model, $\mathcal{M}_{\rm st}$ can be computed explicitly by an ad hoc method, without needing the Riemann surface approach used in the rest of the paper.

For the general 1d random walk, $\det(\lambda\Id-M(g))$ has the form (\ref{det}) with
\begin{eqnarray}
\label{P0 random walk}
&& P_{0}(\lambda)=\tr\Bigg(\prod_{j=1}^{\Omega}\Big(\!\!
\begin{array}{cc}
	\lambda+w_{j+1\leftarrow j}+w_{j\leftarrow j+1} & w_{j\leftarrow j+1}\\
	-w_{j+1\leftarrow j} & 0
\end{array}\!\!\Big)\Bigg)\;,\\
&& P_{+}=-\prod_{j=1}^{\Omega}w_{j+1\leftarrow j}
\qquad\text{and}\qquad
P_{-}=-\prod_{j=1}^{\Omega}w_{j\leftarrow j+1}\;.\nonumber
\end{eqnarray}
For the special case studied in this section, it turns out that the $q$-dependent part of the matrix product above for $P_{0}$ is traceless. Then, $P_{0}(\lambda)$ can be evaluated by taking $q=1$, where all the $2\times2$ matrices are identical, and diagonalization gives
\begin{equation}
\label{P0 almost sym rw}
P_{0}(\lambda)
=\Big(\frac{\sqrt{\lambda}+\sqrt{4+\lambda}}{2}\Big)^{2\Omega}+\Big(\frac{\sqrt{\lambda}-\sqrt{4+\lambda}}{2}\Big)^{2\Omega}\;,
\end{equation}
which is a polynomial in $\lambda$. The only dependency in $q$ of the spectral curve (\ref{spectral curve}) is through
\begin{equation}
P_{\pm}=-(1\pm q)\;.
\end{equation}
Incidentally, this implies from (\ref{det}) that the spectra of $M=M(1)$ and $M(-1)$ are independent of $q$.

The branch points $\lambda_{i}$, solution of $\Delta(\lambda_{i})=0$, are real since the model is a 1d random walk, see section~\ref{section ramification lambda}. When $q\neq0$, the $\lambda_{i}$ are distinct and the hyperelliptic curve (\ref{hyperelliptic curve}) is then non-degenerate. At large $\Omega$, the branch points $\lambda_{i}$ accumulate between $-4$ and $0$ and generate a single branch cut $[-4,0]$ when $\Omega\to\infty$, which appears as a branch cut for $\sqrt{4+\lambda}/\sqrt{\lambda}$ in (\ref{P0 almost sym rw}) but cancels for finite $\Omega$ since $P_{0}$ is a polynomial.

Adaptation of the trace identity (\ref{P0 almost sym rw}) to $M_{\times}$ and $M_{\times}^{\rm R}$ leads for the model considered to
\begin{eqnarray}
\det(\lambda\Id-M_{\times})&=&\frac{\lambda P_{0}'(\lambda)}{\Omega}\\
\det(\lambda\Id-M_{\times}^{\rm R})&=&\frac{\frac{2q^{2}}{\Omega}P_{0}(\lambda)+\big(\frac{1-q^{2}}{\Omega}\,\lambda-\frac{q^{2}}{\Omega^{3}}\,(\lambda+4)\big)P_{0}'(\lambda)}{1-(1-1/\Omega)^{2}q^{2}}\;.
\end{eqnarray}
Additionally, tedious computations lead to $g_{*}=\frac{1}{2}\,\frac{1-q}{1+q}\,P_{0}(\lambda_{*})$ in (\ref{g* zero N}) and $g_{*}^{-1}=\frac{1}{2}\,\frac{1+q}{1-q}\,\big(P_{0}(\lambda_{*})-(1-q^{-1})\lambda_{*}P_{0}'(\lambda_{*})\big)$ in (\ref{gR* zero N}). One can then convince oneself that the poles and zeroes of
\begin{equation}
\label{Mst simple example}
\mathcal{M}_{\rm st}(p)=\frac{q}{\Omega^{2}}\Big(1+\frac{2q}{y}\Big)\,\partial_{\lambda}\frac{2-P_{0}(\lambda)}{\lambda}-\frac{\partial_{\lambda}\big((2-P_{0}(\lambda))^{2}\big)}{2\Omega^{2}\lambda y}\;,
\end{equation}
with $\lambda$ and $y$ evaluated at $p$ after taking the derivative with respect to $\lambda$, match with (\ref{divisor Mst}). Additionally, the behaviour $p\to o$ with average stationary current $J=2q/\Omega^{2}$ matches with (\ref{N(o)}), (\ref{M[N]}). The expression (\ref{Mst simple example}) is thus indeed equal to $\mathcal{M}_{\rm st}(p)$. The time-dependent probability of $Q_{t}$ is then given by (\ref{P[int R]}) with $\mathcal{M}=\mathcal{M}_{\rm st}$, where the closed curve $\Gamma$ can be deformed freely in the complex plane outside the cuts $\gamma_{i}$ since $g^{-1}(0)=\{p_{\infty}^{+}\}$ and $g^{-1}(\infty)=\{p_{\infty}^{-}\}$ for the 1d random walk.

Due to the simplicity of the model, $\mathcal{M}_{\rm st}$ in (\ref{Mst simple example}) is written explicitly in canonical algebraic form (\ref{canonical form mero function}), for arbitrary $\Omega$. Comparison with (\ref{dlogMst}) gives the coefficients $c_{\ell}$. Tedious computations lead to
\begin{eqnarray}
&& \sum_{\ell=1}^{\Omega-1}c_{\ell}\,\lambda^{\ell-1}=\frac{P_{0}(\lambda)-P_{0}(\lambda_{+})+(1+q^{-1})\lambda_{+}\big(P_{0}'(\lambda)-P_{0}'(\lambda_{+})\big)}{4(\lambda-\lambda_{+})}\\
&&\hspace*{35mm} -\frac{P_{0}(\lambda)-P_{0}(\lambda_{-})+(1-q^{-1})\lambda_{-}\big(P_{0}'(\lambda)-P_{0}'(\lambda_{-})\big)}{4(\lambda-\lambda_{-})}\nonumber
\end{eqnarray}
with $\lambda_{\pm}=\frac{4q^{2}}{(\Omega\pm q(\Omega+1))(\Omega\pm q(\Omega-1))}$. One can thus confirm directly in this simple example that the coefficients $c_{\ell}$ are needed in the non-reversible case to ensure that (\ref{dlogMst}) is indeed equal to the differential of the logarithm of a meromorphic function on $\R$. On the other hand, in the reversible limit $q\to0$, one has $\lambda_{\pm}\sim q^{2}\to0$, and the coefficients $c_{\ell}$ do vanish as expected.

\section{Conclusions}
The probability of the process $Q_{t}$ counting the current between two states of a Markov process (and some extensions thereof with the same analytic properties), is given quite generally as the contour integral (\ref{P[int R]}) on a hyperelliptic Riemann surface $\R$ of genus $\gen\geq2$. With stationary initial condition, all the $4\gen$ poles and zeroes of the integrand are known. This leads to the rather explicit formula (\ref{P[int cut] reversible}) for the probability of $Q_{t}$ when the process $Q_{t}$ is reversible. Otherwise, the integrand is given by (\ref{Mst[dlogMst]}), (\ref{dlogMst}), where the $\gen$ constants $c_{\ell}$ have to be fixed so that the integrand is meromorphic on $\R$.

It would be interesting to understand whether the approach considered in this paper can be extended to general initial condition, by characterizing the remaining $\gen$ unknown non-trivial zeroes of the integrand in that case. For stationary initial condition, it would also be nice to identify the $2\gen$ non-trivial zeroes for more general counting processes corresponding to Riemann surfaces that are not hyperelliptic.

Additionally, practical applications of the Riemann surface approach for counting processes would be nice, whether for the analysis of specific models by identifying the relevant meromorphic differentials, or for e.g. disordered models by finding a natural interpretation on the Riemann surface side for the averaging over disorder.

\,\\\noindent\textbf{Funding:} no grant was received for conducting this study.

\,\\\noindent\textbf{Conflicts of interest:} the author has no competing interests that are relevant to the content of this article.

\,\\\noindent\textbf{Data availability:} data sharing is not applicable to this article as no datasets were generated or analysed during the current study.

\vspace{10mm}

\end{document}